\documentclass[twocolumn,final,aps,prx,showpacs,superscriptaddress,amsmath,amssymb,amsfonts,floatfix]{revtex4-1}

\usepackage{graphicx}
\usepackage{multirow}
\usepackage{epsfig}
\usepackage{url}
\usepackage{color}
\usepackage{grffile}

\usepackage[colorlinks,breaklinks,bookmarks=true,citecolor=blue,linkcolor=blue,urlcolor=blue]{hyperref}
\usepackage{color}
\usepackage{tabularx}
\usepackage{sidecap}
\usepackage{soul}
\usepackage{float}

\usepackage[colorlinks]{hyperref}
\usepackage[nameinlink]{cleveref}  
\Crefname{figure}{Fig.}{Figs.}

\newcommand{\ls}[1]{{\color{blue}[LS: #1]}}

\begin{document}

\title{Intertwined charge and spin \textcolor{black}{instability} of La$_3$Ni$_2$O$_7$}

\author{Guiwen Jiang}
\affiliation{School of Physical Science and Technology, Soochow University, Suzhou, China}

\author{Chenye Qin}
\affiliation{School of Physical Science and Technology, Soochow University, Suzhou, China}

\author{Kateryna Foyevtsova}
\affiliation{Department of Physics and Astronomy, University of British Columbia, Vancouver BC, Canada V6T 1Z1}
\affiliation{Stewart Blusson Quantum Matter Institute, University of British Columbia, Vancouver BC, Canada V6T 1Z4}

\author{Liang Si}
\email{siliang@nwu.edu.cn}
\affiliation{School of Physics, Northwest University, Xi'an 710127, China}
\affiliation{Shaanxi Key Laboratory for Theoretical Physics Frontiers, Xi'an 710127, China}
\affiliation{Institute of Solid State Physics, TU Wien, 1040 Vienna, Austria}

\author{Mona Berciu}
\affiliation{Department of Physics and Astronomy, University of British Columbia, Vancouver BC, Canada V6T 1Z1}
\affiliation{Stewart Blusson Quantum Matter Institute, University of British Columbia, Vancouver BC, Canada V6T 1Z4}

\author{George A. Sawatzky}
\email{sawatzky@physics.ubc.ca}
\affiliation{Department of Physics and Astronomy, University of British Columbia, Vancouver BC, Canada V6T 1Z1}
\affiliation{Stewart Blusson Quantum Matter Institute, University of British Columbia, Vancouver BC, Canada V6T 1Z4}

\author{Mi Jiang}
\email{jiangmi@suda.edu.cn}
\affiliation{School of Physical Science and Technology, Soochow University, Suzhou, China}
\affiliation{Jiangsu Key Laboratory of Frontier Material Physics and Devices, Soochow University, Suzhou, China}


\begin{abstract}
  Research on nickel-based superconductors has progressed from infinite-layer LaNiO$_2$ to finite-layer La$_{6}$Ni$_{5}$O$_{12}$, and most recently to the Ruddlesden-Popper phase La$_3$Ni$_2$O$_7$, which was found to exhibits onset of superconductivity at $\sim$80\,K under a pressure of $\sim$16\,GPa.
  Employing density functional calculations and multi-orbital multi-atom cluster exact diagonalization including local exchange and Coulomb interactions, here we analyze the pressure dependent low-energy electronic states of the Ni$_2$O$_9$ cluster, relevant for the bilayer phase of La$_3$Ni$_2$O$_7$. The various possible spin states and the exchange and superexchange mechanisms of the Ni$_2$O$_9$ cluster are quantified via the involvement of the Ni-$3d_{3z^2-r^2}$ orbitals and the atomic Hund's rule exchange, the apical bridging O-$2p_z$ orbitals, and the orbitals involved in the formation of local Zhang-Rice singlet like states. We find that the leading configurations contributiong to the cluster ground-states both for nominal valence and also with local charge fluctuations, do not involve occupation of the apical oxygen, instead they favor formation of in-plane Zhang-Rice singlet like states between an O ligand hole and the Ni $3d_{z^2-y^2}$ orbital.  We also highlight two possible charge and spin ordered states suggested by our cluster results, that are nearly degenerate at all relevant pressures within our modelling. 
\end{abstract}

\maketitle

\section{Introduction}

The discovery of nickelate superconductors \cite{li2019superconductivity,Li2020,PhysRevLett.125.147003,gu2020single} has emerged as a significant milestone in the realm of superconductivity, motivated by the groundbreaking findings of cuprate superconductors since 1986 \cite{Bednorz1986}. Nickelates, with their  layered perovskite structures similar to those of cuprates, have captured the attention of researchers worldwide due to their potential to unveil new insights into unconventional superconducting phenomena \cite{Kitatani2020,PhysRevB.109.235126,Nomura2019,Karp2020,Zhang2019,chen2023electronic,Lechermann2020,PhysRevLett.129.077002,Mi2020,PhysRevB.102.220501}. The origin of superconductivity in nickelates has long been a subject of debate, and the underlying mechanism remains elusive. Recent experiments and theoretical studies have provided compelling evidence suggesting that nickelate superconductivity shares striking similarities with its cuprate counterparts. Specifically, it is believed to be driven by antiferromagnetic (AFM) spin fluctuations and the presence of a nearly half-filled $d_{x^2-y^2}$ band \cite{Kitatani2020} and $d$-wave superconductivity \cite{cheng2024evidence}, reminiscent of the cuprate scenario \cite{PhysRevB.109.235126}.

The recent discovery of superconductivity of La$_3$Ni$_2$O$_7$ 
\cite{sun2023signatures,PhysRevX.14.011040,yang2024orbital,hou2023emergence,zhou2024evidence,PhysRevLett.132.256503,jiang2024high,wang2024normal,dan2024spindensitywave,wu2024superexchange,xie2024neutron,yi2024antiferromagnetic}
has added a new dimension to the landscape of nickelate superconductors. This compound introduces a novel archetype of nickelate superconductors, characterized by the presence of apical oxygen between Ni cations with an average valence of Ni$^{2.5+}$ as compared to Ni$^{3+}$ (3$d^7$) in the cubic NdNiO$_3$ or Ni$^{1+}$ (3$d^9$) in the infinite-layer NdNiO$_2$. The existence of more than one 3$d$ hole per Ni introduces a new complication by  involving other 3$d$ orbitals such as the $d_{3z^2-r^2}$ orbital and its strong coupling via Hund's rule exchange with the $d_{x^2-y^2}$ orbital of the same Ni. These external factors are believed to activate the Ni-3$d_{z^2}$ and inner apical O-2$p$ orbitals besides Ni-3$d_{x^2-y^2}$ \cite{PhysRevB.108.L201121,wang2024electronic,PhysRevB.108.125105}, thus influencing the electronic structure and superconducting properties of the material.

Regarding the superconducting mechanism of La$_3$Ni$_2$O$_7$, recent studies have explored several key areas. Firstly, there is ongoing debate surrounding the geometrical crystal structure, with recent works reporting the coexistence of ``2222" (bilayer) and ``1313" (monolayer+trilayer) ordering of the NiO$_2$ layers \cite{cui2024strain,abadi2024electronic,chen2024polymorphism,wang2024long,zhang_high-temperature_2024,La2PrNi2O7_2024}. 
Secondly, apical O vacancies have been reported to mainly occupy the inner apical positions and strongly influence the electronic and magnetic structure~\cite{dong_visualization_2024}. This would drive the Ni to preferentially take +2 valence, {\em i.e.} 4 holes to reside on the bilayer Ni dimer. For example, the compound La$_3$Ni$_2$O$_{6.5}$ has half of the apical O missing and all Ni take $d^8$ configurations~\cite{Cava24}. 
Thirdly, controversies persist regarding the electronic structure and active orbitals involved in the superconducting mechanism of La$_3$Ni$_2$O$_7$ \cite{PhysRevB.108.214522,PhysRevB.109.L180502,PhysRevB.108.L201121,wang2024electronic,PhysRevB.108.125105}. Current investigations into potential superconducting mechanisms include the exploration of charge-lattice coupling arising from the breathing mode \cite{PhysRevMaterials.2.125001,chen2023critical,zhan2024cooperation}, superconducting instability attributed to multiple orbitals and magnetic exchange interactions, the emergence of a Cu-based superconducting structure due to $d$-$p$ hybridization under pressure, and the role of magnetic exchange interactions resulting from strong occupation of both $d_{x^2-y^2}$ and $d_{3z^2-r^2}$ atomic orbitals. 
\textcolor{black}{The pairing symmetry has also been extensively investigated. A dominant $s$-wave pairing has been suggested, with electron-phonon coupling or isotropic interactions playing a central role \cite{PhysRevB.108.L140505,PhysRevB.109.L220506,luo2024high,ouyang2024absence,PhysRevB.109.104508,PhysRevLett.132.146002,PhysRevB.108.165141,PhysRevB.109.165154,PhysRevLett.133.096002,PhysRevLett.132.036502,PhysRevB.108.174501,PhysRevLett.133.136001,PhysRevB.109.L180502,PhysRevB.108.L140504,PhysRevB.108.214522}. In contrast, $d$-wave pairing, driven by strong electronic correlations, has also emerged as a candidate, supported by theoretical frameworks commonly used to describe other oxides superconductors \cite{jiang2024high,wu2024superexchange,PhysRevB.109.104508,PhysRevLett.132.146002,PhysRevB.109.L180502,PhysRevB.110.L060510,PhysRevB.108.214522}. 
Some studies proposed spin-density wave (SDW) order as a competing or coexisting phase, where fluctuations associated with SDW order mediate the pairing \cite{geisler2024optical,zhang2024doping,chen2024electronic,PhysRevMaterials.8.L111801,kakoi2024multiband,PhysRevLett.131.206501,PhysRevB.110.195135,PhysRevB.110.L140508,PhysRevB.108.125105,PhysRevLett.132.256503}. 
Antiferromagnetic (AFM) spin fluctuations, a hallmark of many correlated transition-metal oxides, have been widely discussed as a possible driver for the observed superconductivity \cite{PhysRevB.108.L201121,chen2024electronic,shen2023effective}.}
Additionally, electron-phonon coupling in La$_3$Ni$_2$O$_7$ was investigated and found insufficient to drive the high $T_c$ superconductivity \cite{ouyang2024absencephononmediatedsuperconductivityla3ni2o7}. 
This is why it remains important to combine various methods to include both the correlation effects and local magnetic moments on Ni.

In this work, we aim to delve into the electronic properties of the nickelate superconductor La$_3$Ni$_2$O$_7$ using a combination of  Density-Functional Theory (DFT) \cite{PhysRev.136.B864,PhysRev.140.A1133} and local cluster models that include local electron correlations and multiplet structures. By employing these computational techniques, we seek to elucidate the ground-state  and low-energy eigenstates, as well as their orbital occupation with and without doping.  \textcolor{black}{The nature of these leading configurations provides valuable clues about the low-energy physics of these bilayers, and its similarities and differences from cuprate physics.  Our results suggest two possible types of intertwinned charge density combined with spin density wave states (CDW-SDW) orders that are nearly degenerate at ambient pressure, in good agreement with results of DFT simulations~\cite{Kateryna}.}

\textcolor{black}{The paper is organized as follows: Section II presents the model and our methods, Section III presents our results and their analysis, and Section IV contains a summary and outlook.}

\section{Model and Method}

\subsection{DFT calculations and Wannier projections}

The DFT level structural relaxations and electronic band structure calculations were performed using the \textsc{Vasp}  \cite{kresse1996efficiency,PhysRevB.54.11169} and \textsc{Wien2K} \cite{blaha2001wien2k,Schwarz2002} code with the Perdew-Burke-Ernzerhof version of the generalized gradient approximation (GGA-PBE) \cite{PhysRevLett.77.3865} and a dense $k$-mesh for different phases of La$_3$Ni$_2$O$_7$, including $Cmmm$, $Amam$, $Fmmm$ and $I4/mmm$. Specifically, for $Cmmm$, $Amam$ and $Fmmm$ phase the $k$-mesh is set at 12$\times$12$\times$3, while it is 13$\times$13$\times$3 for the $I4/mmm$ phase. 
To obtain the hopping parameters between Ni-$d$ and O-$p$ orbitals, the Ni-$d$ and O-$p$ bands from DFT (\textsc{Wien2K}) calculations around the Fermi energy are projected onto Wannier functions \cite{PhysRev.52.191} using Wannier90~\cite{mostofi2008wannier90,RevModPhys.84.1419} and the \textsc{Wien2Wannier} \cite{kunevs2010wien2wannier} interface. 
In DFT calculations and Wannier projections we found that the magnitude of the electron hopping values are mainly determined by the applied pressure, with the tilting and rotations of the octahedrons playing a relatively smaller role, hence, the electron hopping parameters of the $I4/mmm$ phase at 0, 4, 8 and 16\,GPa, and $Fmmm$ phase at 29.5\,GPa are shown in Table~\ref{table}.

\subsection{Multi-orbital cluster model}

\begin{figure}[t!]
\psfig{figure=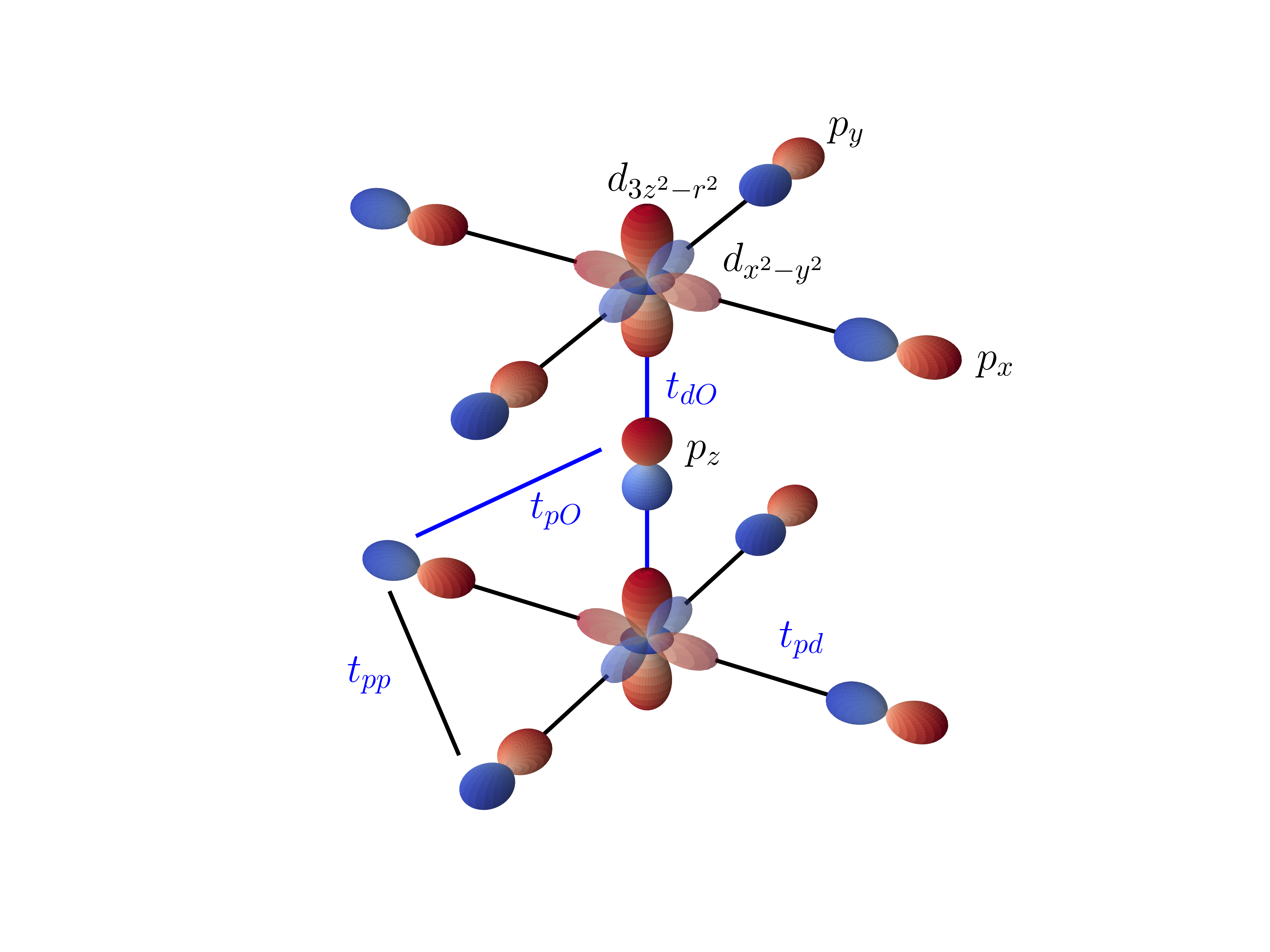,width=.5\textwidth, trim={3cm 0.5cm 0.6cm 0.9cm}, clip}
\caption{Schematic geometry of the Ni$_2$O$_9$ cluster studied in this work, showing the O and Ni orbitals kept in the basis of configurations, as well as the Ni-O and O-O hopping processes included in the model.  }
\label{geom}
\end{figure}

We consider a bilayer NiO$_2$ lattice with two Ni ions sitting at the center of each layer, sandwiching an additional interlayer (apical) Oxygen. This cluster is approximated as an isolated molecule with an average of 5 holes in otherwise filled Ni-$3d$ and O-$2p$ valence orbital shells as displayed in Fig.~\ref{geom}. This approach is motivated by our previous simulations on the single layer model with one single Ni ion relevant to the infinite-layer nickelates~\cite{Mi2020,Mi2020a,Mi2022,Mi2023,PhysRevB.108.155147}, where we found that the local physics dominates. As a result, we expect that the current Ni$_2$O$_9$ cluster, i.e., two Ni ions and their four nearest neighbor O in plane plus one shared apical nearest neighbor O as depicted in Fig.~\ref{geom}, will provide valuable information on the intrinsic physics of La$_3$Ni$_2$O$_7$. Another practical reason for this cluster calculation is that going beyond this, even to a double Ni$_2$O$_9$ cluster with a total of  4 Ni ions and 16 O and a total of 10 holes, would necessitate further approximations to the number of configurations assumed to contribute to the low energy physics.  
Nevertheless, below we  will \textcolor{black}{use these single-cluster results to speculate on possible magnetic and charge orders in this material,   thus connecting with the various magnetic and electronic structure studies in hybrid functional calculations.}
To this aim, we will consider two neighboring Ni$_2$O$_9$ clusters with 6 hole and 4 hole separately, which represent the electron removal (photoemission) and electron addition (inverse photoemission) states. 

The average Ni valence, if we assume that O is always $-2$, would be $+2.5$. 
However, we note that even in the case of the cubic NdNiO$_3$ with formal Ni$^{3+}$ valence, the Ni is dominantly in a 2+ valence $d^8$ configuration, with one hole per formula unit in the O-$2p$ orbitals, {\em ie.}  2 holes on average on the octahedron around each Ni~\cite{bisogni_ground-state_2016,PhysRevLett.112.106404,PhysRevB.94.195127}.
For La$_3$Ni$_2$O$_7$, because of its bilayer structure and the equally short Ni to interlayer apical O bond lengths as compared to the Ni to in-plane O bond length, the fifth hole besides the four on the two  Ni 3d$^{8}$ could be housed in either the interlayer or in-plane O.
Hence, there is a potential competition between placing the 5th hole in the plane as in the case of the hole-doped cuprates, and placing it in the apical O between the NiO$_2$ layers. 
One of the important questions we address here is how the distribution of 5 holes in the 2222 structure of La$_3$Ni$_2$O$_7$ containing the cluster of Fig.~\ref{geom} varies with different parameters, particularly the pressure which affects interatomic distances and the O-2p to Ni-3d hybridization. 

We conducted the cluster exact diagonalization calculation as in our previous works~\cite{Mi2020,Mi2020a,Mi2022,Mi2023,PhysRevB.108.155147} to investigate the nature of parent La$_3$Ni$_2$O$_7$.
The general Hamiltonian reads as:
\begin{equation}\label{Ham} {\cal H} =
\hat{U}_{dd}+\hat{U}_{pp}+\hat{T}_{pd}+\hat{T}_{pp}+\hat{T}_{dO}+\hat{T}_{pO}+\hat{E}_s
\end{equation}
where $\hat{U}_{dd}$ includes all Coulomb and exchange interaction of the $3d^8$ multiplet corresponding to $D_{4h}$ symmetry in terms of Racah parameters $A,B,C$, which are linear combinations of conventional Slater integrals. $\hat{U}_{pp}$ denotes the onsite interaction of Oxygen ligand 2$p$ orbitals.
$ \hat{T}_{pd}$, $\hat{T}_{pp}$, $\hat{T}_{dO}$, $\hat{T}_{pO}$ incorporate hopping integrals between the inter-layer Oxygen labeled as $O$ (only including the most relevant $p_z$ orbital) and the Ni-$3d_{z^2}$ orbitals and in-plane O-$2p$ ligand orbitals. Conventionally, $L$ denotes the linear combination of four ligand O orbitals nearest to a Ni with a particular symmetry. For example, the combination of $x^2-y^2$ and $x^2+y^2$ symmetries hybridize with $3d_{x^2-y^2}$ and $3d_{3z^2-r^2}$ orbitals separately. 
Finally, $\hat{E}_s$ describes the on-site energies of various Ni-$3d$, in-plane and interlayer O-$2p$ states. 
 

We determine the nature of the ground state (GS) as well as the excited states of the multihole systems, and  the weights of the various configurations' contributions to the ground states. We focus on the dependence on a variety of parameters such as $t_{dO}$ and $t_{pd}$, which have the strongest influence as a function of pressure.

To label the multiple-hole states, we use notation such as $d^8$-$O$-$d^9$ to denote the configuration where the top Ni layer, interlayer Oxygen, and bottom Ni layer are occupied as $d^8$, $O$, and $d^9$ states separately.
Note that all configurations with asymmetric components in the two layers, such as $d^8$-$O$-$d^9$, have a mirrored configuration via inversion symmetry between layers;  their linear combinations result in bonding ($+$) and antibonding ($-$) states $(d^8$-$O$-$d^9\pm d^9$-$O$-$d^8)/\sqrt{2}$. The interlayer O only hybridizes with the antibonding state of two Ni-$d_{z^2}$ orbitals via $t_{dO}$ owing to $3d_{z^2}$ and apical $2p_z$ orbitals' phases.
Furthermore, in detailed labeling like $\{d_{z^2}d_{x^2}\} \{d_{z^2}d_{x^2}\}(S=0)$, we use  $\{ . \}$ and $[ . ]$ to denote the two spins forming triplet and singlet states, respectively, with the shorthand notation $d_{x^2}\equiv d_{x^2-y^2}$ and $d_{z^2} \equiv d_{3z^2-r^2}=d_{2z^2-(x^2+y^2)}$ respectively. 
$S$ gives the total GS spin of the cluster.

\begin{table}[t!]\label{table}
\footnotesize
\caption{On-site energies $\epsilon$, Racah parameters $A,B,C$, and hopping integrals $T^{pd}_{mn}$ with $m \in \{d_{x^2}, d_{z^2}\}$ with $d_{x^2} \equiv d_{x^2-y^{2}}$ and $d_{z^2} \equiv d_{3z^2-r^2}=d_{2z^2-(x^2+y^2)}$ respectively and $n \in \{p_x, p_y \}$, where $m,n$ are nearest neighbors, extracted from DFT calculations. Note that we only consider $p_x$ and $p_y$ orbitals for in-plane O denoted by $p$ with lobes pointing to the Ni ion; while only the $p_z$ orbital is considered for the  interlayer apical Oxygen denoted by $O$. 
Only the magnitudes of hopping integrals are shown, while the sign is determined by the overlap of the lobes of different orbitals. The DFT values of $t_{dO}$ and $t_{pd}$ are only for reference and they will be varied as two major control parameters throughout the work. All values are in units of eV.}

\centering
\begin{tabular}{c|c|c| c c c | c c} 
 \hline\hline

  $T^{pd}_{x^2-y^{2},n}$  & $T^{pd}_{z^2,n}$  &
  $\epsilon(d_{x^2})$ & $A$ & $B$ &  $C$  & $U_{OO}$ & $U_{pp}$ \\ [0.5ex] 
 \hline
 $t_{pd}$ &  $t_{pd}/\sqrt{3}$  & 0.0 & 6.0 &  0.15 & 0.58 & 4.0 & 4.0  \\ 
 \hline 
\end{tabular} 

\centering
\begin{tabular}{c | c c c c c | c c c c} 
 \hline
    &
  $\epsilon(d_{z^2})$  & $\epsilon(d_{xy})$  &
  $\epsilon(d_{xz/yz})$  & $\epsilon_p$ &  $\epsilon_O$ 
  & $t_{pd}$ & $t_{pp}$ &  $t_{dO}$  & $t_{pO}$\\ [0.5ex] 
\hline
0 GPa &  0.046 & 0.823 & 0.706 & 2.47  & 2.94 & 1.38 &  0.537 & 1.48 & 0.445\\ [0.5ex] 
\hline
4 GPa &  0.054 & 0.879 & 0.761 & 2.56  & 3.03 & 1.43 &  0.548 & 1.53 & 0.458\\ [0.5ex] 
\hline
8 GPa &  0.060 & 0.920 & 0.804 & 2.62  & 3.02 & 1.46 &  0.554 & 1.55 & 0.468\\ [0.5ex] 
 \hline
16 GPa &  0.072 & 0.997 & 0.887 & 2.75  & 3.14 & 1.52 &  0.566 & 1.61 & 0.484\\ [0.5ex] 
 \hline
29.5 GPa &  0.095 & 1.06 & 0.94 & 2.9  & 3.24 & 1.58 &  0.562 & 1.66 & 0.487\\ [0.5ex] 
 \hline\hline
\end{tabular} 
\label{table}
\end{table}

The parameters listed in Table~\ref{table} give the on-site energies of the two $e_g$ orbitals, in-plane and interlayer O-$2p$ orbitals limited to their $\sigma$ bonded to the Ni-$3d$ orbitals, as well as the hopping integrals between various orbitals. 
We emphasize that we are limiting the basis set to the $e_g$ orbitals as an approximation since, in the case of the cubic perovskites with even 3 holes per Ni formally, the D$_{4h}$ crystal and ligand field splitting dominate over the multiplet interactions to put the 3 holes in $e_g$ orbitals.
These parameters are determined from the DFT \cite{PhysRev.136.B864} calculation and Wannier \cite{PhysRev.52.191,mostofi2008wannier90} projections using \textsc{WIEN2k} \cite{blaha2001wien2k,Schwarz2002} and \textsc{WIEN2WANNIER} \cite{kunevs2010wien2wannier} and revised Perdew-Burke-Ernzerhof for solids (PBESol) of the generalized gradient approximation (GGA) \cite{PhysRevLett.100.136406} for the treatment of exchange-correlations functional. 
As usual, the site energy of Ni-$3d_{x^2-y^2}$ is set to be zero as reference.
The two most crucial parameters are the hybridization $t_{pd}$ and $t_{dO}$ illustrated in Fig.~\ref{geom}, which are adopted as two control parameters to  mimic the experimental pressure effects.

\begin{figure*}[t]
\psfig{figure=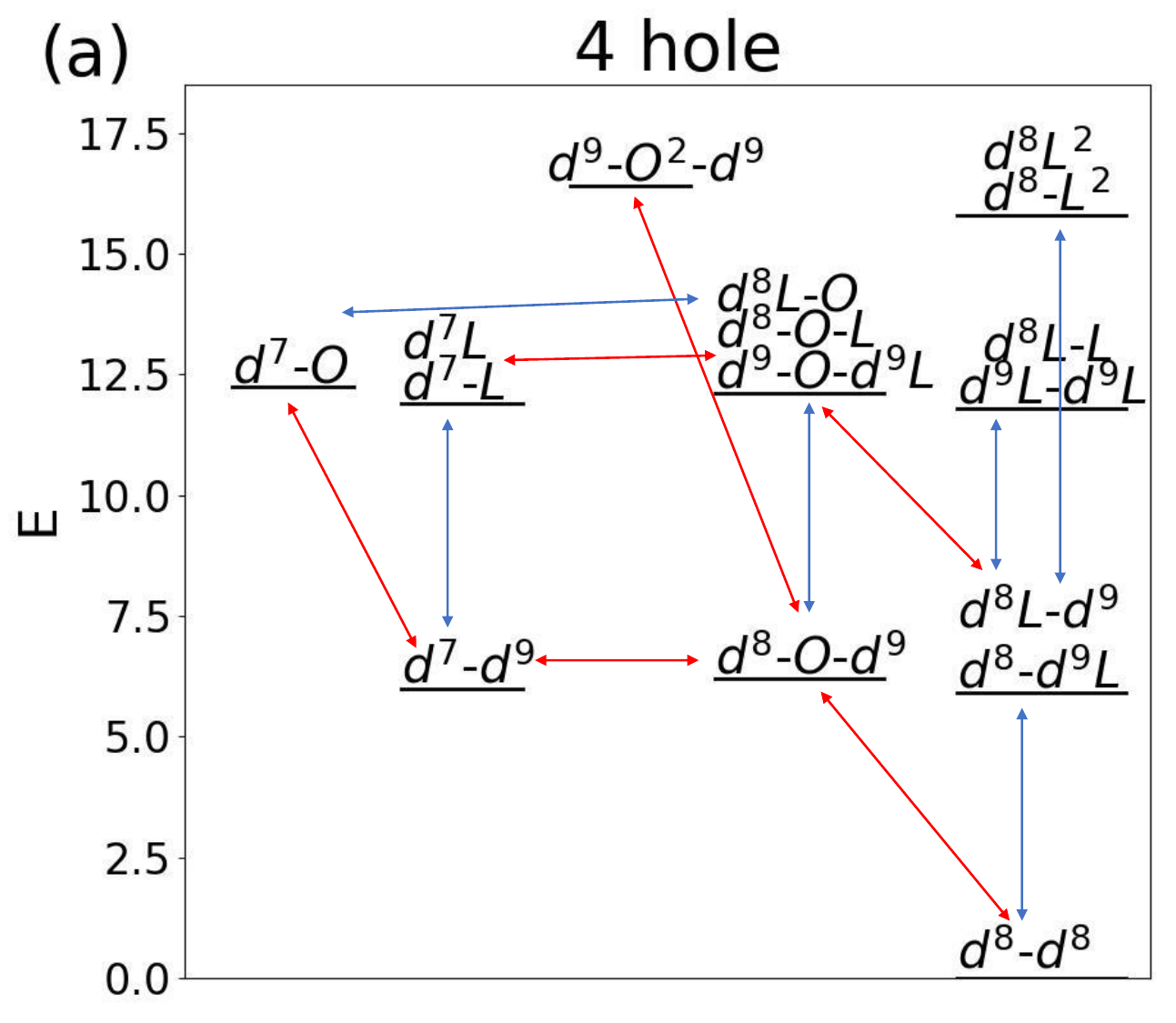,width=.32\textwidth, clip}
\psfig{figure=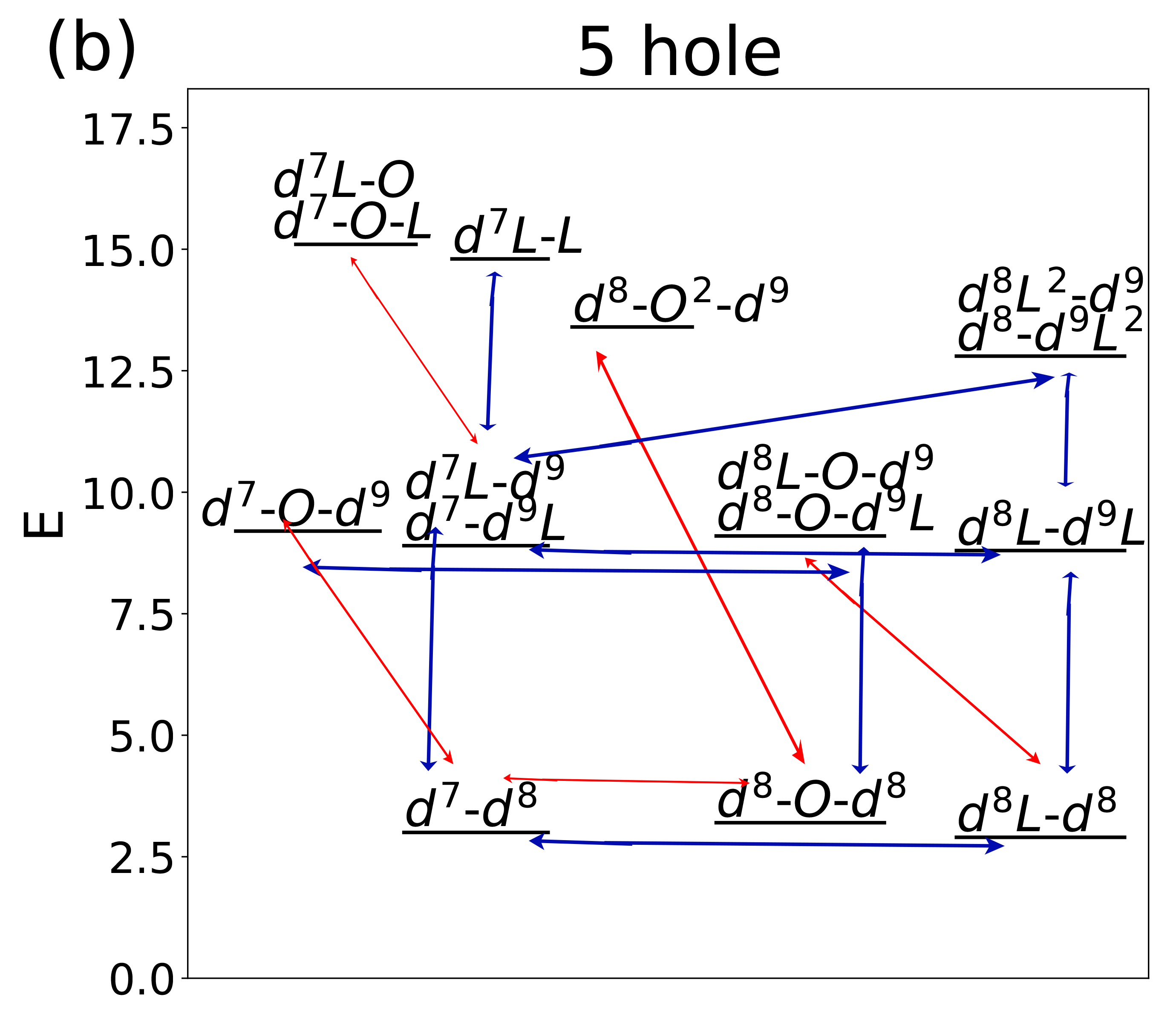,width=.32\textwidth, clip}
\psfig{figure=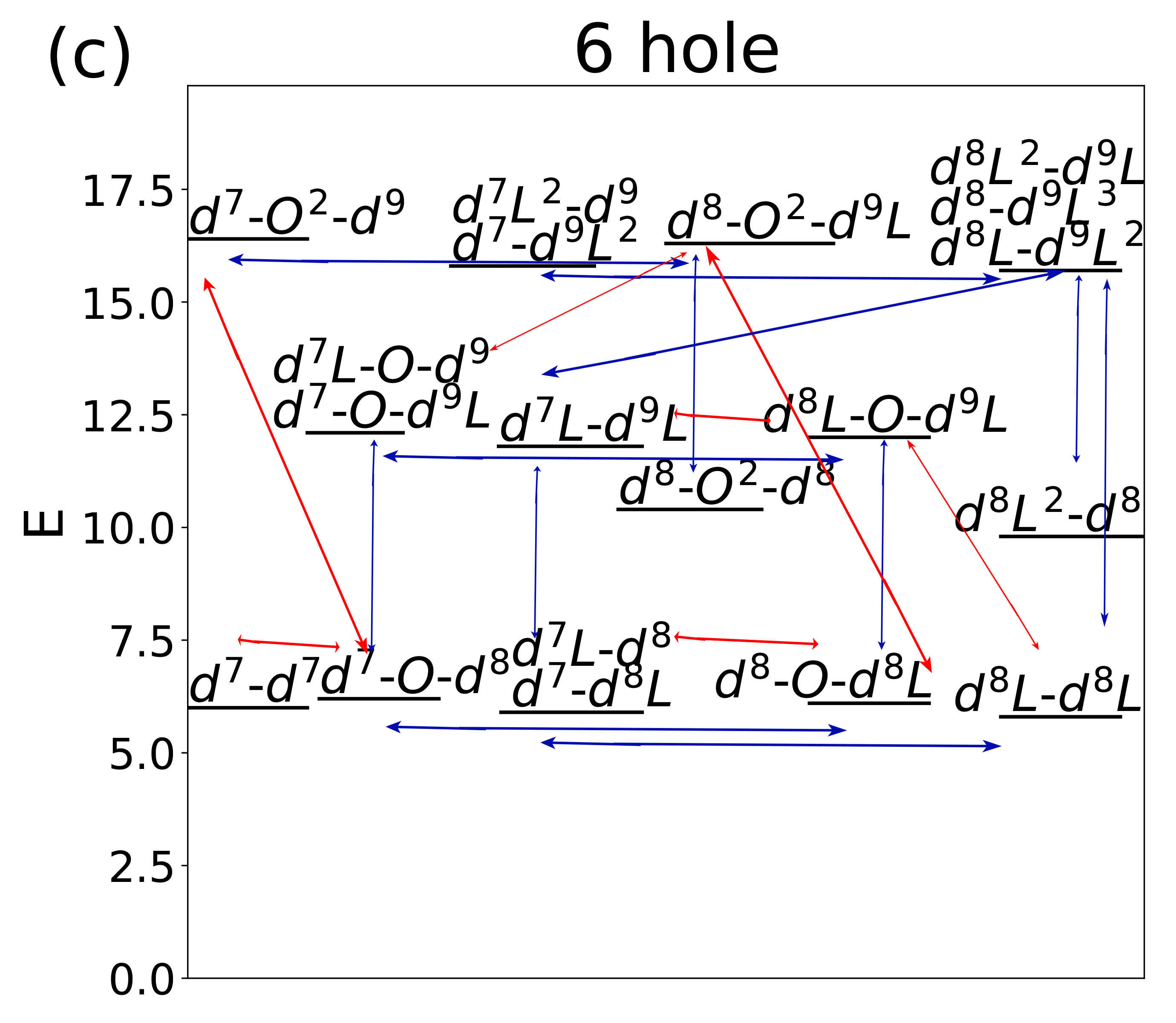,width=.32\textwidth, clip}
\caption{Energy level diagrams of 4, 5, 6 hole clusters (from left to right) \textcolor{black}{in the absence of hybridization and correlations apart from the Ni $U$}. The red (blue) arrow indicate $t_{dO}$ ($t_{pd}$) hybridization matrix elements. 
The 4-hole $d^8$-$d^8$ is chosen as the zero energy configuration. 
These configurations are highly degenerate and this degeneracy will be lifted when switching on the Hund's rule exchange and various hopping integrals.
Configurations with even higher energies are not shown for clarity.}
\label{ediagram}
\end{figure*}

The conventional Racah parameters $A=6.0$ eV, $B=0.15$ eV, $C=0.58$ eV  
 describe the on-site Coulomb and exchange interactions of the $3d$ electrons on Ni~\cite{Mi2020}. The parameter $A$ is close to what one generally refers to as the Hubbard $U$ in the nickelates. The $B$ and $C$ parameters, obtained from atomic physics, describe the difference in the energies of two holes in Ni $3d$ and their dependence on the spin and orbital relative orientation. 
Additionally, the Hubbard interaction  describing two holes on the same O has been measured using Auger spectroscopy for oxides~\cite{PhysRevB.38.11322}, which finds  a value of 4.6\,eV for Cu$_2$O (full shells are needed to do this measurement properly with Auger). Because Ni is somewhat smaller, we adopt the estimated 4\,eV for both apical and in-plane O's onsite Hubbard interaction. 

The vacuum state is defined as $3d^{10} 2p^6$, {\em i.e.}  fully occupied Ni-$3d$ and O-$2p$ orbitals for both in-plane and interlayer Oxygen. To differentiate the in-plane from the interlayer Oxygen, whose site energies may differ from each other in the high pressure superconducting phase, from now on we denote them by L and O, respectively.
Owing to the Ni$^{+2.5}$ valence, there are on average 5 holes per Ni$_2$O$_9$ cluster in the undoped parent compound, therefore  we focus on the 5 hole states' spin, orbital, and position distribution. Additionally, we will also explore the 4-hole and 6-hole states, as mentioned above. 


\subsection{Energy level diagram}

Before proceeding, we emphasize some important considerations on our strategy of treating the 5-hole problem, especially the choice of the zero energy reference states, so as to have some initial understanding of the ground states in the absence of any hybridizations.

The lowest energy 5-holes eigenstate can be regarded as either the lowest energy electron removal state from the 4-hole ground state, or the lowest energy electron addition state of the 6-hole ground states. 
For the 4-hole system, the 4 holes will be evenly distributed into two layers by the inversion symmetry of the system so that the 4-hole leading GS configuration is expected to be $d^8$-$d^8$ due to the relatively large charge transfer energy compared to the cuprates~\cite{PhysRevLett.126.127401} and also the stabilization of the triplet ($S=1)$ ground state of Ni $d^8$ due to Hund's rule exchange between the two holes on $d_{x^2-y^2}$ and $d_{z^2}$ orbitals.
As we will show next, this is consistent with what we find in the 4-hole calculation once we turn on the hybridizations, as the ground state that is strongly separated in energy from all other configurations.

Recall that the effective Ni Hubbard $U$ is defined as the energy difference between electron removal and electron addition from $d^8$. In our previous work on the infinite-layer nickelates~\cite{Mi2020}, the dominant ground state configuration of Ni was $d^9$, so we took it as the reference state and place the $d^{10}$ and $d^8$ states at energies $U/2$ above $d^9$, giving the correct energy separation $U=E(d^8)+E(d^{10})-2E(d^9)$.      
For our 4-hole system relevant to this cluster and considering $d^8$ as the starting zero energy state, it is appropriate to separate the Hubbard $U$ evenly onto the $d^7$ and $d^9$ states, so that their energies both lie at $U/2$ above the $d^8$ state.

To have a full understanding of all configurations, Fig.~\ref{ediagram} illustrates the energy level diagrams in the absence of hybridizations. 
Note that these configurations are highly degenerate involving different multiplets. This degeneracy will be lifted when switching on the Hund's rule exchange and the hopping integrals which introduce additional exchange interactions such as the ligand and O hole's AFM exchange with  Ni-$d$ holes, which is a large interaction of up to several eV, as we will see below. The smaller superexchange terms coupling the upper and lower Ni-$d_{z^2}$ spins  vary between 10-200\,meV and will also be discussed below. 
For example, the 6-hole configurations can have a total spin  $S=0,1,2,3$ and their degeneracy will be lifted due to the exchange interactions mentioned above.
Here, with the Racah parameter $A$ denoting the interaction strength akin to Hubbard $U$ in single-orbital models, we have $E(d^8)=0$, $E(d^7)=E(d^9)=A/2=3$ eV. The blue and red double arrows denote the $t_{dO}$ and $t_{pd}$ hybridizations respectively. 
These involve energy scales up to several eV and will result in large exchange interactions and energy splitting of the nearly degenerate lowest energy states especially for the cases of 5 and 6 holes. For example, in the case of 5 holes, the states involving the hole on $O$ or $L$  are nearly degenerate, but this will be strongly lifted because of the effectively strong hybridization involving the $L$ hole and also the much stronger $d_{x^2-y^2}$-$L_{x^2-y^2}$ exchange interaction. As a result, for parameters close to the DFT values, the states with a hole on the apical $O$ are much higher in energy than those with instead an in-plane  $L$ hole of $x^2-y^2$ symmetry, as we will show below.

\section{Results and discussion}

In what follows, we describe in detail the results of the exact diagonalization for the 4, 5, and 6 hole Ni$_2$O$_9$ clusters in terms of the various configurations shown in Fig.~\ref{ediagram} but now with all the interactions switched on, including the effects of hybridization caused by $t_{pd}$, $t_{dO}$, and $t_{pp}$. We consider the results  for an extended range of these parameters. Figures ~\ref{4hole}--\ref{6holedE} provide information on the  ground state energy  of the 4, 5, 6 hole systems in terms of the contributions of the various configurations and spins to the ground state wave functions. We focus on the physically relevant situations with fixed ratio $t_{dO}/t_{pd} \sim 1.05$ motivated by the DFT parameters. The red star indicates the location of the DFT parameter values. The notations used for the description of the dominant states provides detailed information as to which spin and orbital occupations are involved.

\subsection{4-hole system}

\begin{figure}[t]
\includegraphics[width=0.45\textwidth]{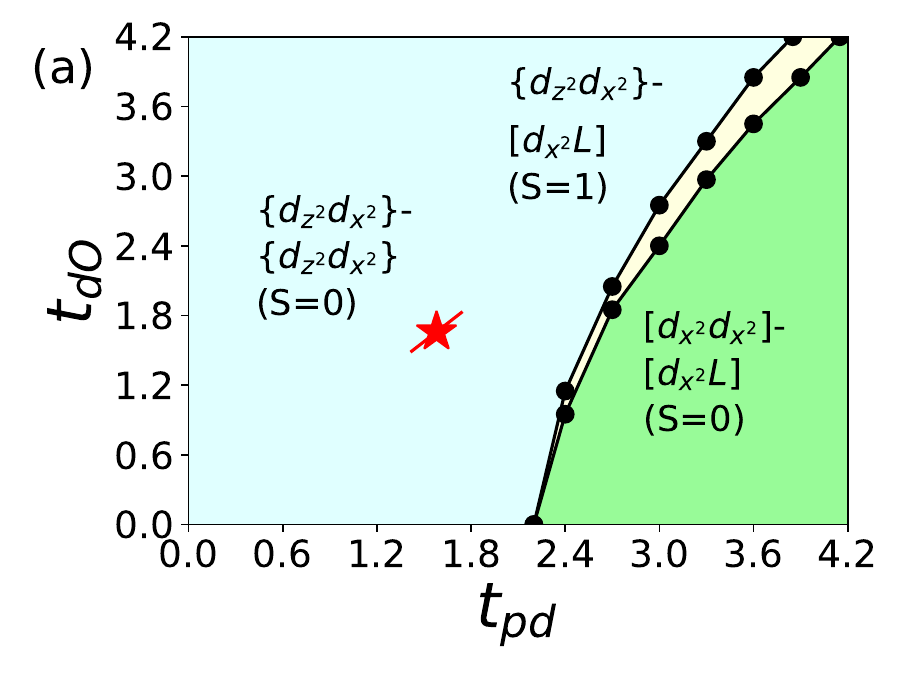}
\includegraphics[width=0.45\textwidth]{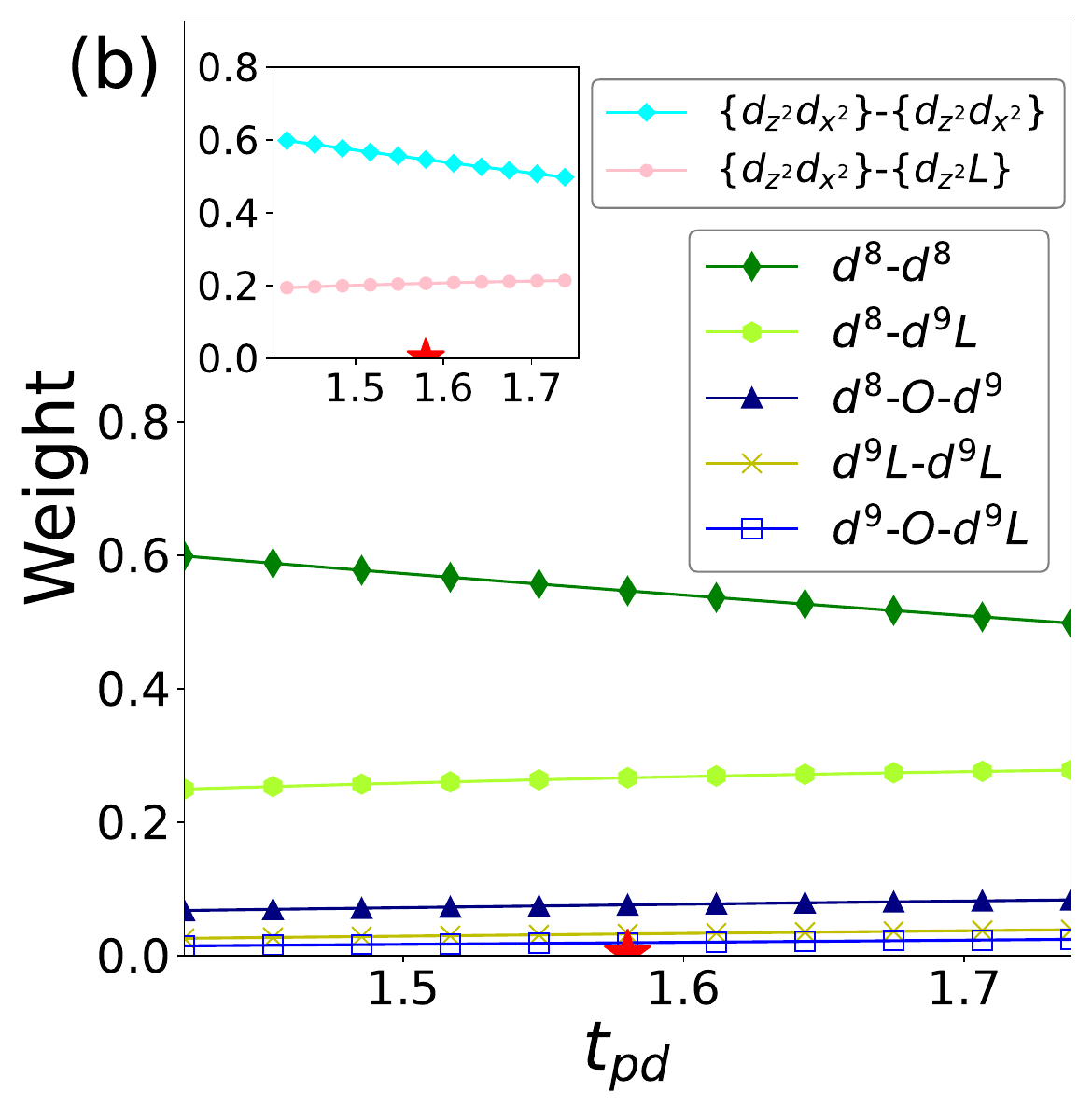}
\caption{\textcolor{black}{(a) Ground-state phase diagram of a 4-hole cluster as a function of $t_{pd}$ and $t_{dO}$. The star shows the high pressure 29.5 GPa DFT parameters from Table~\ref{table}; the red line indicates evolution with pressure of these DFT hopping integrals,  with a fixed $t_{dO}/t_{pd}=1.05$. For all realistic values, the 4-hole ground-state has a total spin $S=0$, achieved through AFM coupling of the two Ni $d^8$, both of which are in their $s=1$ triplet state due to the strong Hunds coupling. As discussed,  $\{ . \}$ and $[ . ]$ denote two hole spins  forming triplet and singlet states, respectively.(b) Evolution with increasing pressure, {\em i.e.} along the red line indicated in panel (a),  of the GS weights of the various listed configurations. The inset provides more details  on the leading configurations. }
}
\label{4hole}
\end{figure}

Figure~\ref{4hole}(a)  illustrates \textcolor{black}{the GS phase diagram of the  the 4-hole system as a function of $t_{pd}$ and $t_{dO}$. As mentioned, the red star shows the DFT parameters relevant for high-pressure  superconducting La$_3$Ni$_2$O$_7$, and the red line shows the  fixed ratio $t_{dO}/t_{pd}=1.05$, consistent with the DFT values when the pressure is varied. Figure~\ref{4hole}(b) shows the evolution  of the GS weights of the most important configurations as pressure is increased, {\em i.e.} for hopping values along the red line shown in panel (a).  }  Clearly, for a wide range of parameters, the GS is predominantly characterized by a total spin $S=0$ configuration composed of two in-plane $\{d_{z^2}d_{x^2}\}$ triplet states that are antiferromagnetically coupled to each other, confirming the expectation from the analysis of the energy diagram Fig.~\ref{ediagram}(a).  

Panel (b) shows that an increasing  $t_{pd}$ gradually suppresses the weight of this  dominant $d^8d^8$ configuration promoting $d^8d^9L$ instead, which corresponds to an in-plane Zhang-Rice singlet (ZRS). 
The inset further reveals the detailed spin and orbital decomposition of the dominant configurations. Generically, the 4-hole system is the playground of competition between Hund's rule preferred $\{d_{x^2-y^2}d_{z^2}\}$ triplet and the ZRS with an admixture of the low spin $[d_{x^2-y^2}L_{x^2-y^2}]$ configuration via large $t_{pd}$ and ligand field splitting between $d_{x^2-y^2}$ and $d_{z^2}$ orbitals.

\begin{figure}[h!]
\centering
\includegraphics[width=0.44\textwidth]{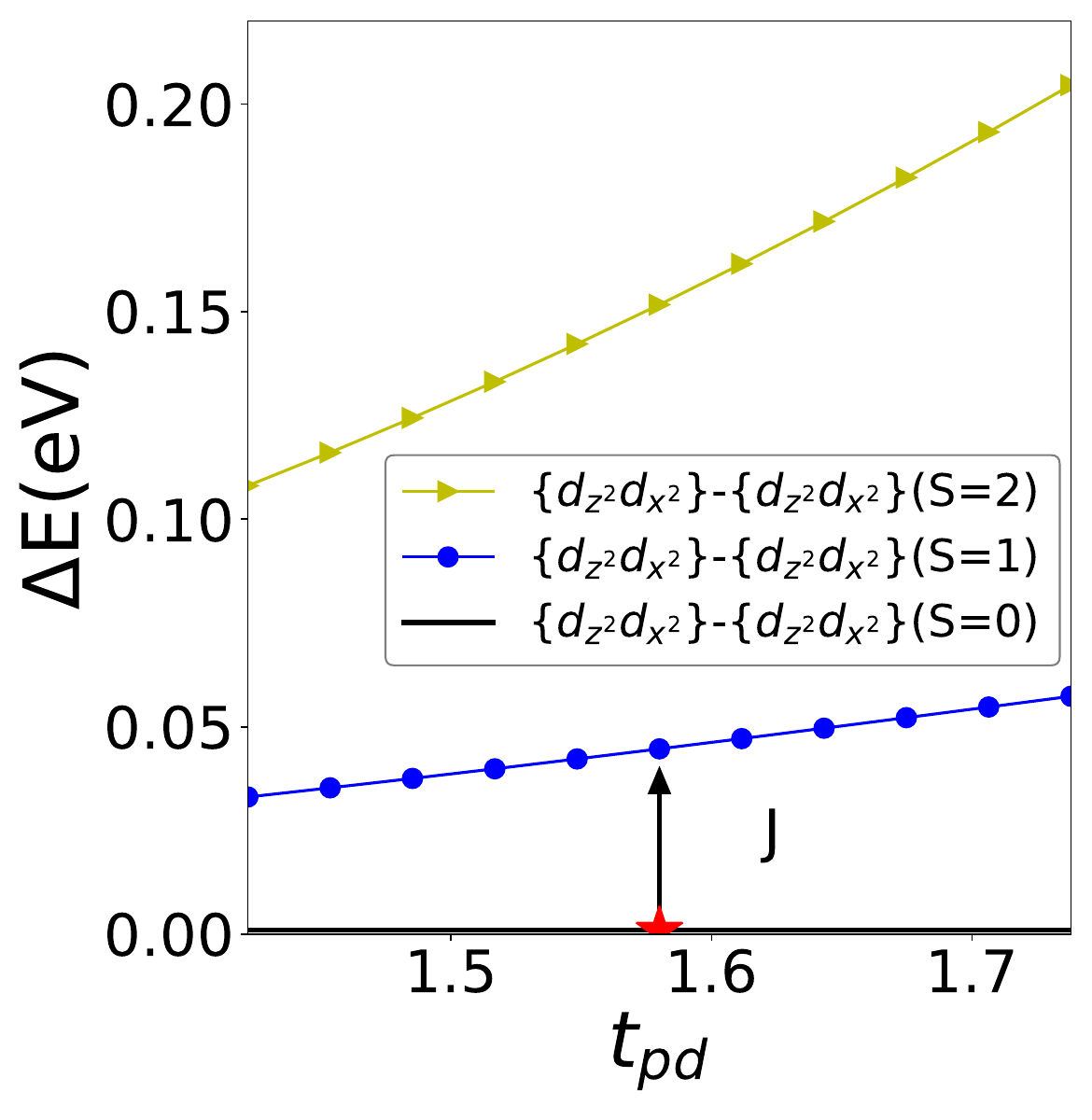}
\caption{Energy difference between 4-hole lowest-energy spin $S=1$ (circles) and $S=2$ (triangles) eigenstates and the GS with $S=0$, for fixed $t_{dO} / t_{pd} = 1.05$. $J$ labels the superexchange interaction energy at the high pressure 29.5 GPa DFT parameters.
}
\label{4holedE}
\end{figure}

In addition to these orbital occupations, it is also interesting to look at the lowest-energy eigenstates with higher total spin, as these may be important in considering the interactions between the Ni$_2$O$_9$ clusters via in plane hopping of ligand holes. \textcolor{black}{The two $s=1$ Ni spins are coupled through an AFM superexchange $J_O$ mediated by the apical oxygen, leading to total spin states of $S=0,1,2$, among which the total spin of 2 would involve the ferromagnetic alignment of the two spin-1 $d^8$ triplets, while the spin-0 state is their antiferromagnetic alignment. (Of course, there are many other much higher-energy eigenstates involving, for instance,  $s=0$  singlets at one or both Ni sites. These are not shown.)} The variation  with $t_{pd}$, upon applying pressure, of the energy differences between these  low-lying total spin states are shown in Fig.~\ref{4holedE}. 
Since the total spin $S$ is a good quantum number with eigenvalue $S(S+1)= s_1(s_1+1) +s_2(s_2+1) + 2\vec{s}_1 \cdot \vec{s}_2$ where $ \vec{s}_1$ and $ \vec{s}_2$  describe the $s_1=s_2=1$ triplet $d^8$ on the two Ni, the low-energy spin only Hamiltonian is $J_O\vec{s}_1 \cdot \vec{s}_2$ with superexchange $J_O$. Indeed, we see that the energy splitting between $S=0,1$ is about half of that between $S=1,2$. The deviations are a result of the \textcolor{black}{spin dependent changes in the spatial part of the wave functions for large $t_{pd}$ and $t_{dO}$, which is an important part of the whole discussion when considering the very high energy states. Nevertheless, from these results we estimate that $J_O$ increases from about 50\,meV to about 300\,meV with increasing pressure}. 
This small energy scale is important because it may be smaller than inter-cluster interactions.

\subsection{5-hole system}

\begin{figure}[t]
\includegraphics[width=0.45\textwidth]{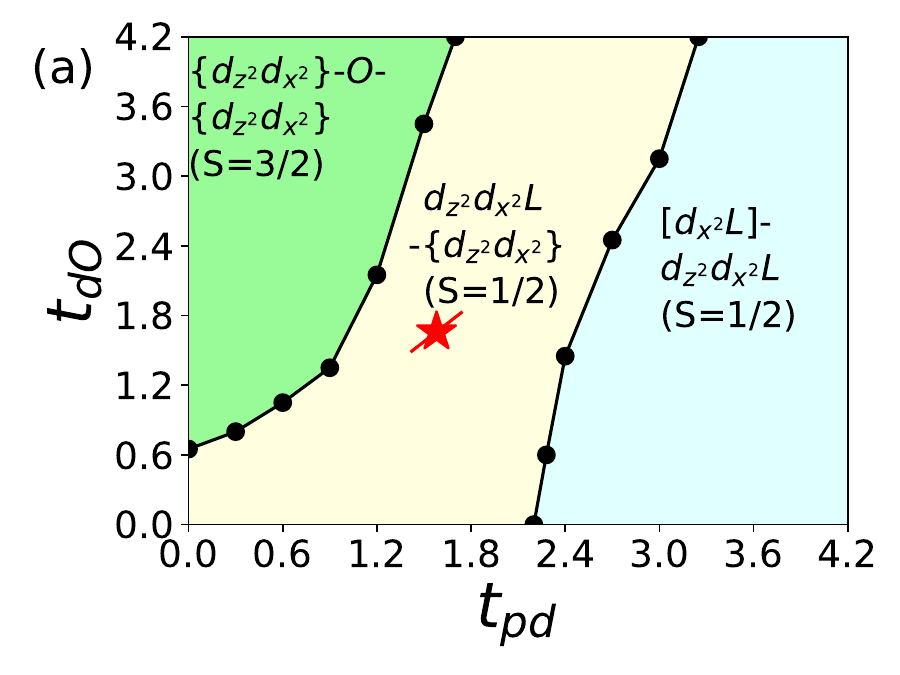}
\includegraphics[width=0.45\textwidth]{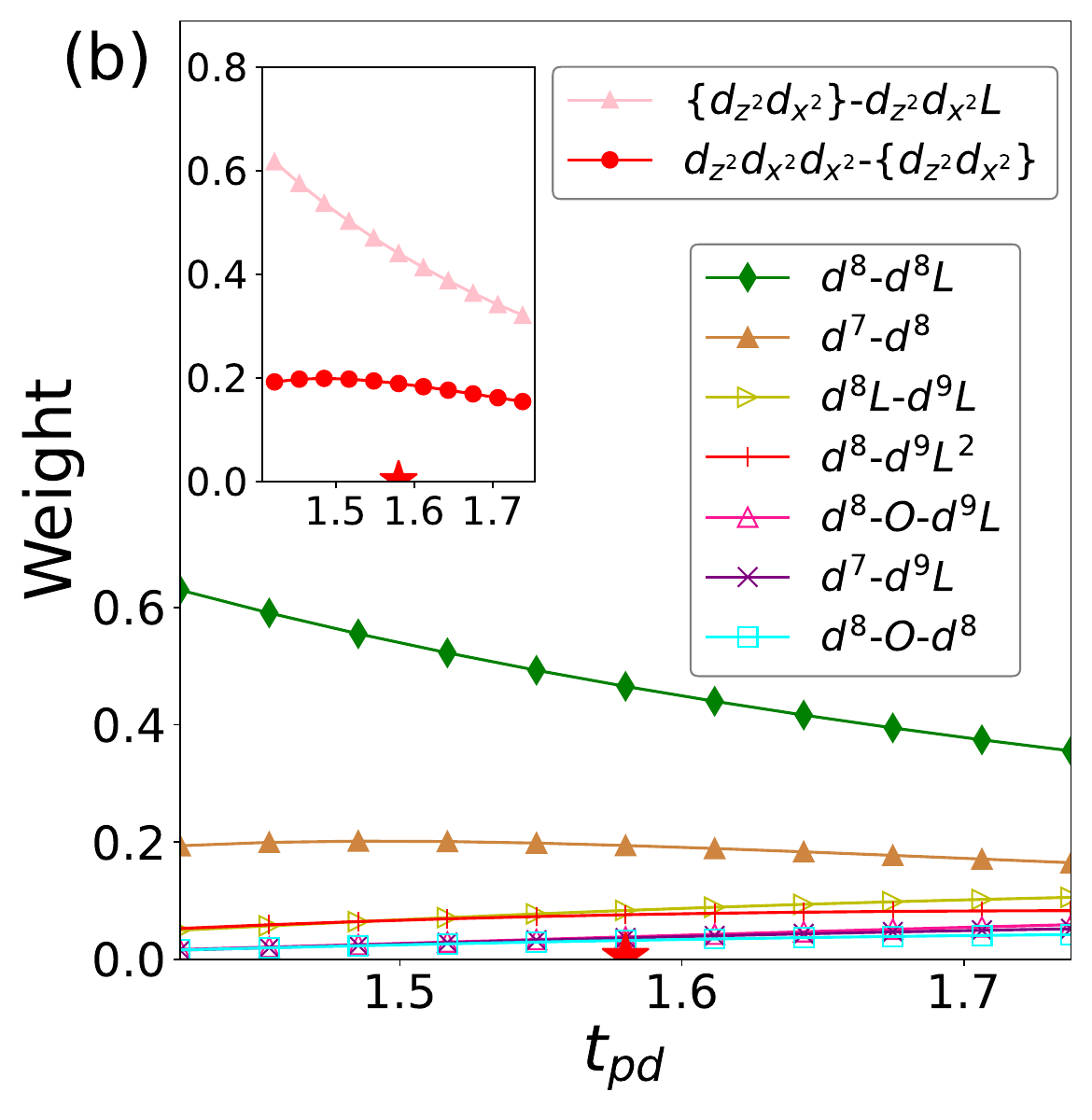}
\caption{Same as Fig.~\ref{4hole} but for a  5-hole cluster. Again, the red star indicates the DFT predicted parameters  at  high pressure of 29.5 GPa. and the red line is their variation as the pressure is varied. Panel (b) shows the GS weights of the most important configurations along this line.
}
\label{5hole}
\end{figure}

Now we switch to the 5-hole system \textcolor{black}{corresponding to the average hole concentration in undoped La$_3$Ni$_2$O$_7$}. Figure~\ref{5hole} shows the phase diagram and GS weight distribution akin to Fig.~\ref{4hole}, with the red star denoting the DFT parameters at high pressure. 
\textcolor{black}{We start by noting that configurations with a hole on the apical O between the NiO$_2$ planes only contribute  strongly to the GS for parameters far from the DFT values, with  small $t_{pd}$ and large $t_{dO}$ (left side of panel (a)). In this region of the parameters space, the ground state has a total spin 3/2, followed by states with $S=1/2$ and 5/2 at higher energies. We consider this region to be unphysical based on the values of the DFT parameters. 
For DFT predicted parameters (central region in panel (a)), the 5-hole GS is dominantly of $d^8Ld^8$ character, {\em i.e.} the 5th hole is  doped onto the in-plane Oxygen starting from the previously discussed 4-hole GS. This is consistent with the expectations based on the analysis of the energy level diagram in Fig.~\ref{ediagram}(b). }

This $L$ hole state has $x^2-y^2$ symmetry and forms a Zhang-Rice like singlet (well known in hole-doped cuprates) with the $d_{x^2-y^2}$ hole. This state is strongly stabilized by the large exchange interaction between these two spins, which is much larger than the Hund's rule exchange. This leaves a spin-1/2 on the  $d_{3z^2-r^2}$ orbital in this layer, and a spin-1 $d^8$ triplet state for the Ni in the other layer. Together, they combine to form the total spin $S=1/2$ ground state of the Ni$_2$O$_9$ cluster in the parameter region close to the DFT based parameters. We note that this $L$ hole can reside in either the upper or lower NiO$_2$ plane, and  that hopping of the $L$ hole between the NiO$_2$ planes is not allowed in our DFT based model Hamiltonian.  

As shown in Fig.~\ref{5hole}(b), while the configuration $d^8Ld^8$ discussed above has the highest weight to the GS for DFT parameters,  the first subleading configuration is $d^7d^8$ and is followed by at least 5 other  configurations that also contribute significantly to the ground state wave function. These configurations  vary by  one and two hole changes from the dominant $d^8Ld^8$ state; this is a clear indication that strong correlations are involved.

\begin{figure}[t]
\includegraphics[width=0.44\textwidth]{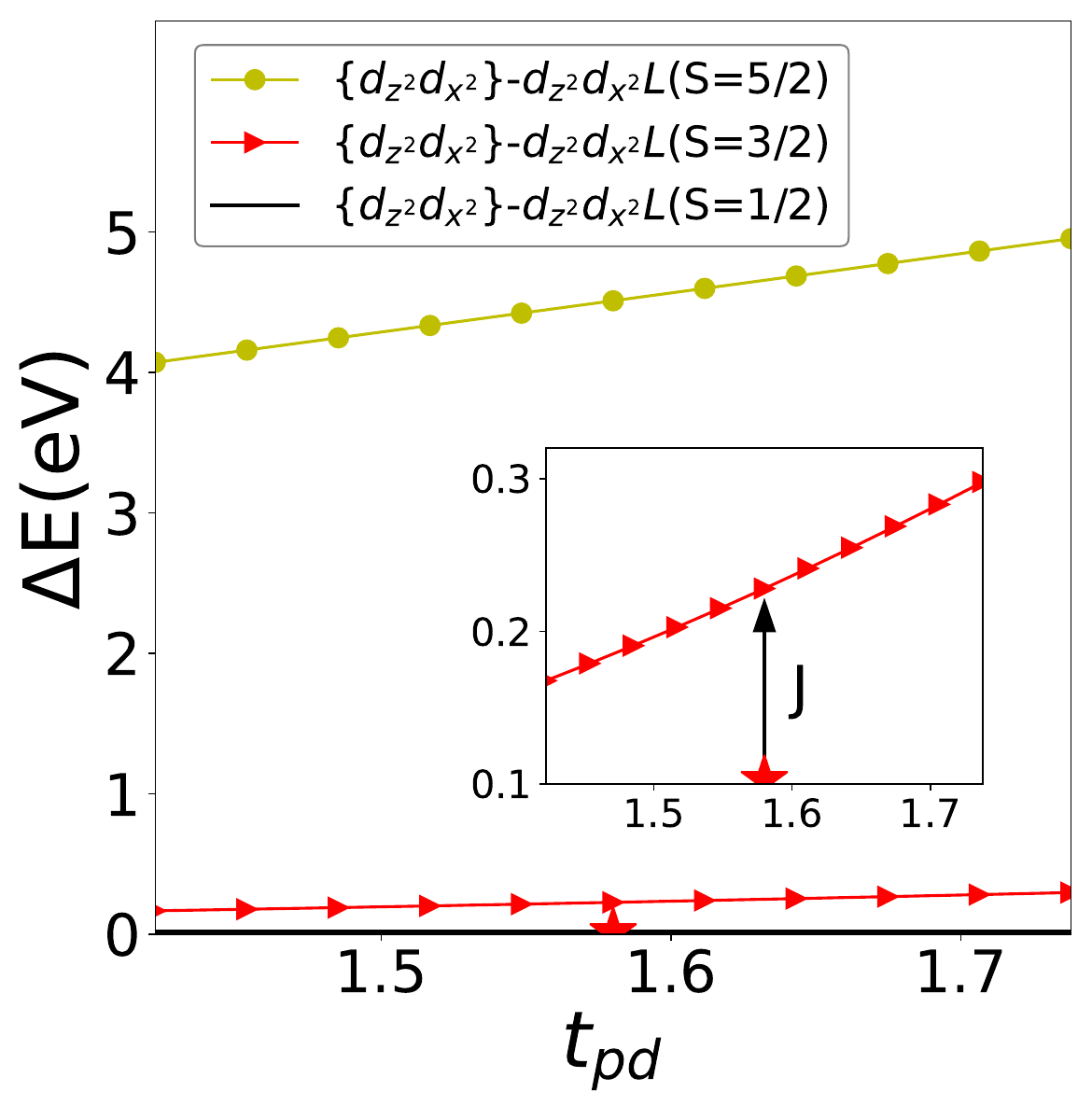}
\caption{Energy difference between 5-hole lowest-energy spin $S=3/2$ (triangles) and $S=5/2$ (circles) eigenstates and the GS with $S=1/2$, for fixed $t_{dO} / t_{pd} = 1.05$. $J$ labels the superexchange interaction energy at the high pressure 29.5 GPa DFT parameters.
}
\label{5holedE}
\end{figure}

\begin{figure}[t]
\psfig{figure=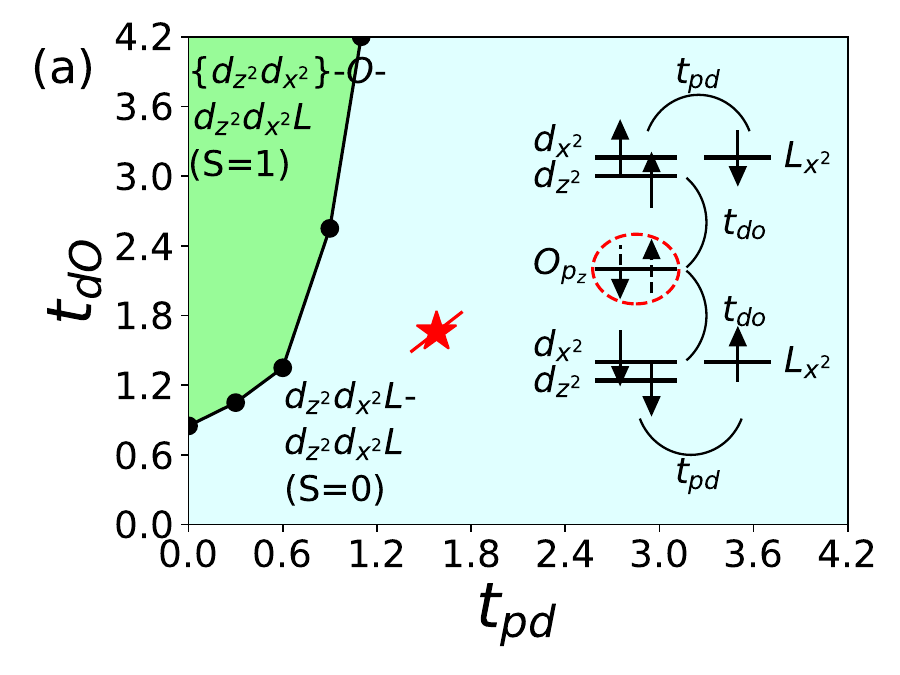,width=.45\textwidth, clip}
\psfig{figure=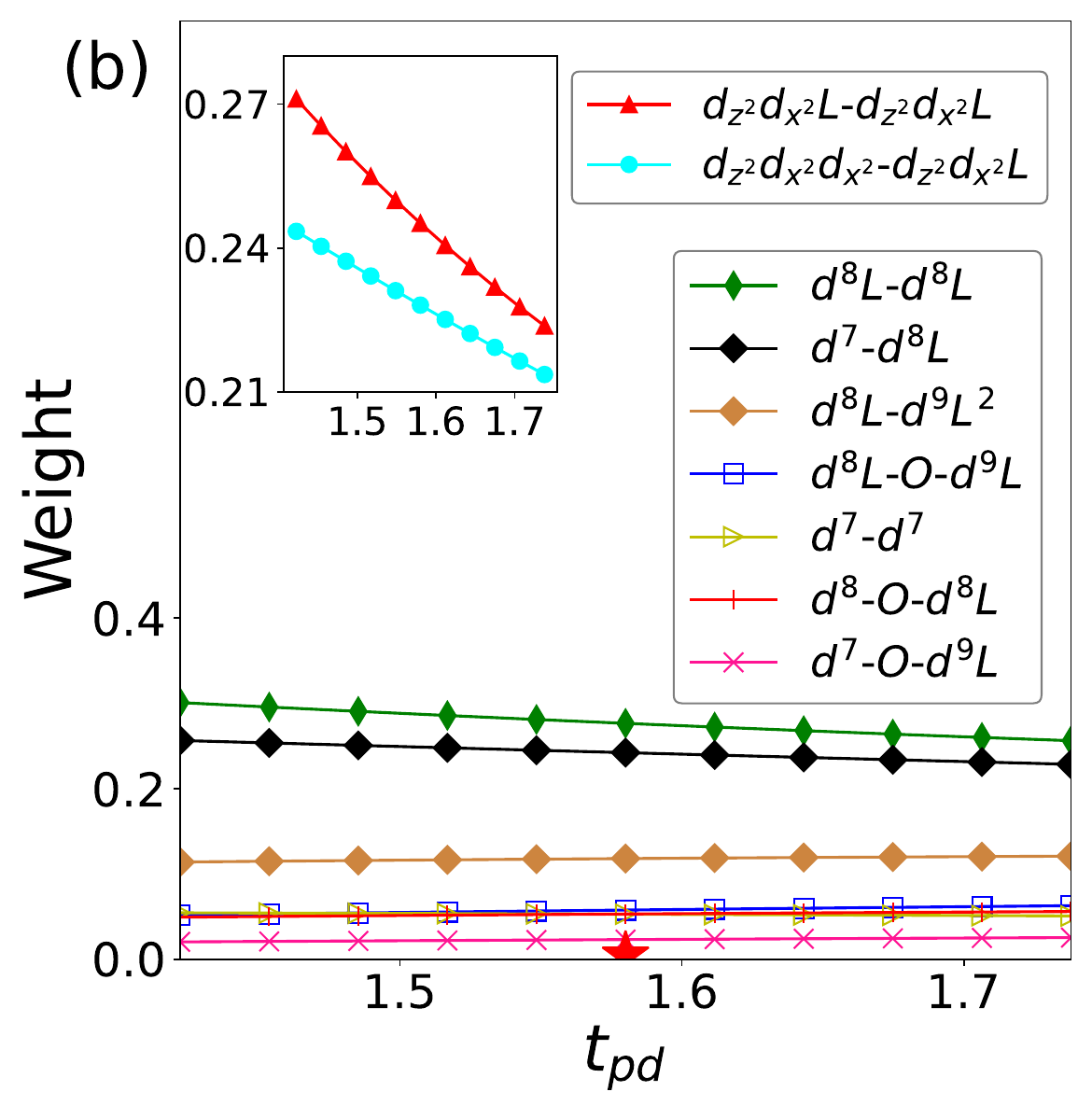,width=.45\textwidth, clip}
\caption{Same as Fig.~\ref{4hole} and Fig.~\ref{5hole} but for a 6-hole cluster.
}
\label{6hole}
\end{figure}

These strong correlations are most obvious when looking at the lowest energy states with possible total spin $S=1/2$, $3/2$, $5/2$. As discussed, the $S=1/2$ state is the preferred ground state for parameters close to the DFT values. Nonetheless, other spin states may become important when considering the inter-cluster interactions. Fig.~\ref{5holedE} shows the energies of the lowest-energy $S=3/2$, $5/2$ states relative to the $S=1/2$ GS state (black line as reference). The exchange interaction due to the antiferromagnetic superexchange coupling via interlayer O is about 200\,meV at high pressure (shown in the inset) with the $S=5/2$ state at extremely high energy  since it requires all 5 spins to be parallel, implying  a ZR triplet rather than singlet state stabilized via the aforementioned huge exchange between the $L_{x^2-y^2}$ and $d_{x^2-y^2}$ holes~\cite{PhysRevB.41.288}. \textcolor{black}{The very high-energy of this $S=5/2$ state is beyond a simple perturbative approach, and the splitting cannot be explained with a simple Heisenberg-type Hamiltonian, because the many-body wavefunction that generates this state is quite different from that of the states with $S=3/2, 1/2$. However, in a simple perturbative approach the exchange splitting between the ZRS and its triplet counterpart is  $4t^2_{pd}/\Delta\approx 3.4$eV for DFT values. It is this large energy scale that is primarily responsible for the strong stabilization of the ZR singlet state involving the $L_{x^2-y^2}$ and $d_{x^2-y^22}$ orbitals}. 


\subsection{6-hole system}

We illustrate the 6-hole's GS phase diagram and weight evolution with $t_{pd}$ in Fig.~\ref{6hole}.
Firstly, for the DFT parameters the GS has low spin $S=0$, akin to the 4-hole system. The $t_{pd}$ hybridization favors having the holes in the two layers forming the leading $d^8L$-$d^8L$ and the subleading $d^7$-$d^8L$ configurations. 
As in the 5 hole system, the $d^8L$ configuration is strongly stabilized due to the huge exchange interaction between the $L_{x^2-y^2}$ hole and the $d_{x^2-y^2}$ hole, favouring the formation of a ZRS and leaving a spin-1/2 in each of the $d_{3z^2-r^2}$ orbitals. These, in turn, are antiferromagnetically coupled due to the superexchange via the apical $O$. 
The importance of the subleading $d^7$-$d^8L$  configuration is also clear, as its weight is only slightly smaller than that of the dominant $d^8L$-$d^8L$ configuration with inversion symmetry.

\begin{figure}[t!]
\centering
\psfig{figure=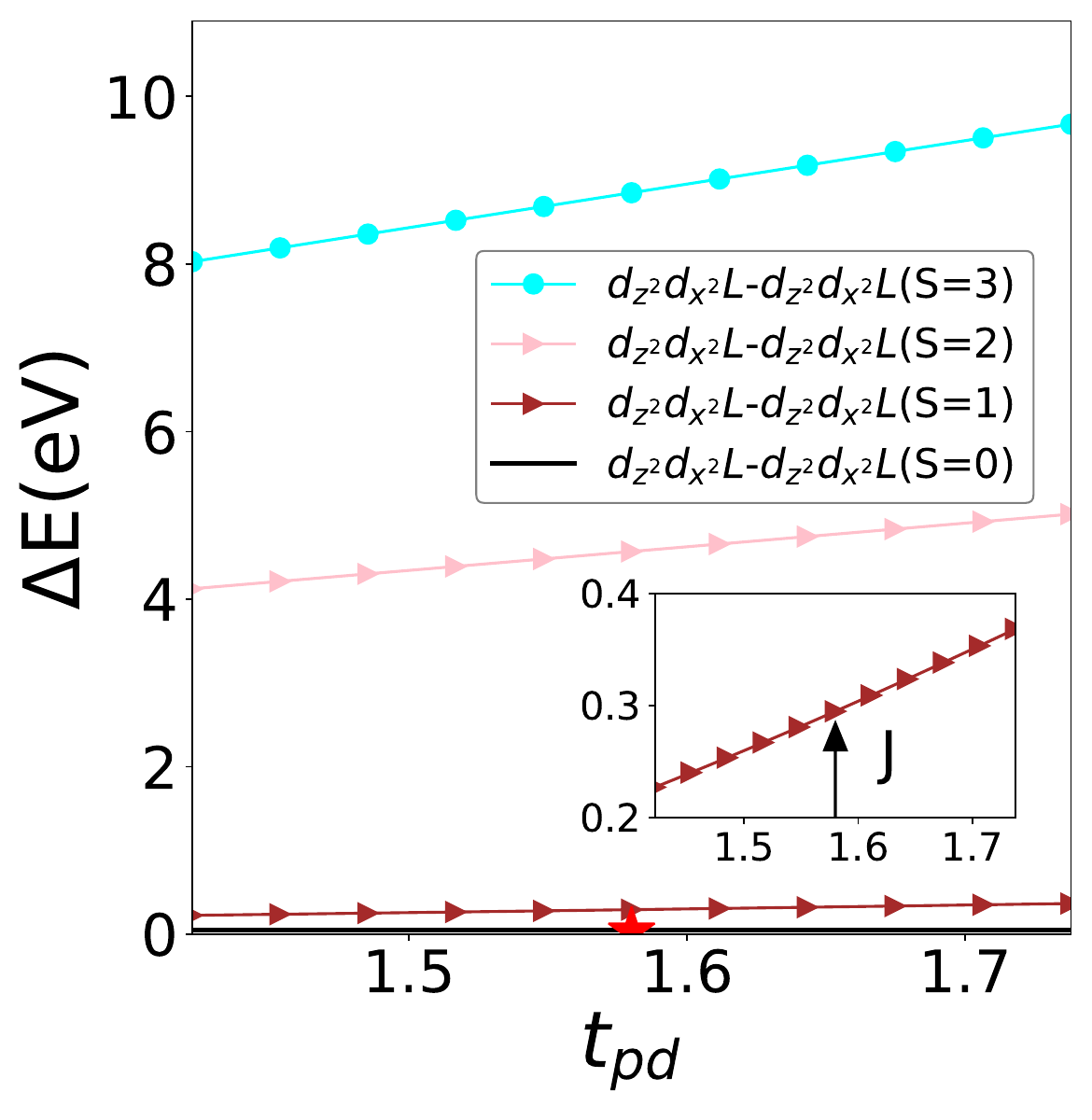,width=.44\textwidth, clip}
\caption{Energy difference between 6-hole lowest-energy spin $S=1,2,3$ eigenstates and the GS with $S=0$, for fixed $t_{dO} / t_{pd} = 1.05$. $J$ labels the superexchange interaction energy at the high pressure 29.5 GPa DFT parameters.
}
\label{6holedE}
\end{figure}

Although the ground state of the 6 hole cluster has $S=0$, it is again important to consider the excited higher-spin states since they could be significant when considering the inter-cluster interactions. In Fig.~\ref{6holedE}, we show the parameter dependence of the energies of the $S=1,2,3$ spin states relative to the lowest energy $S=0$ state. We see that the lowest-energy $S=1$ state is at about 300\,meV above the GS and may be of importance when considering the inter-cluster interactions. We also note that its energy is very weakly dependent on $t_{pd}$, distinct from the states with $S=2,3$. The $S=2$ state has to involve the triplet Zhang-Rice like state, which is about 4\,eV above the GS,  reminiscent of the 5-hole cluster. In addition, the maximum $S=3$ state must involve all spins parallel so both the $d^8L$ in both layers must involve the ZR triplets, explaining its much higher energy (around 8 or 9\,eV above $S=0$ for the high pressure DFT parameters). \textcolor{black}{Similar to the 5-holes cluster, these very high energy scales for the high-spin states involving the ZR triplets are not well described in an exchange only, Heiseberg-type model, because the radial part of the wave functions involved are very different from those of the $S=0,1$ states.} 
These high energies are far beyond any possible influence of the inter-cluster interactions so we can safely neglect them. 




\subsection{\textcolor{black}{Potential charge  and spin density wave states}}

Next, we explore potential ordered states by considering the dominant orbital and spin configurations of the lowest energy states in the 4-, 5-, and 6-hole Ni$_2$O$_9$ clusters. 
To summarize the results so far, we found that the 4-hole GS is dominated by the configuration $d^8(s=1)$-$d^8(s=1)$ with total spin of $S=0$, separated by an energy of about 50\,meV from its $S=1$ counterpart. The 5-hole ground state is dominated by $d^8L(s=1/2)$-$d^8(s=1)$ degenerate with the $d^8$-$d^8L$ configuration, and  has a GS total spin $S=1/2$ and the lowest-energy excited state with $S=3/2$ at about 200\,meV above it. Finally, the 6 hole state is dominated by the $d^8L(s=1/2)$-$d^8L(s=1/2)$ configuration with a GS total spin $S=0$ and the first excited $S=1$ state at about 300\,meV higher energy.

\begin{figure}[t]
\centering
\psfig{figure=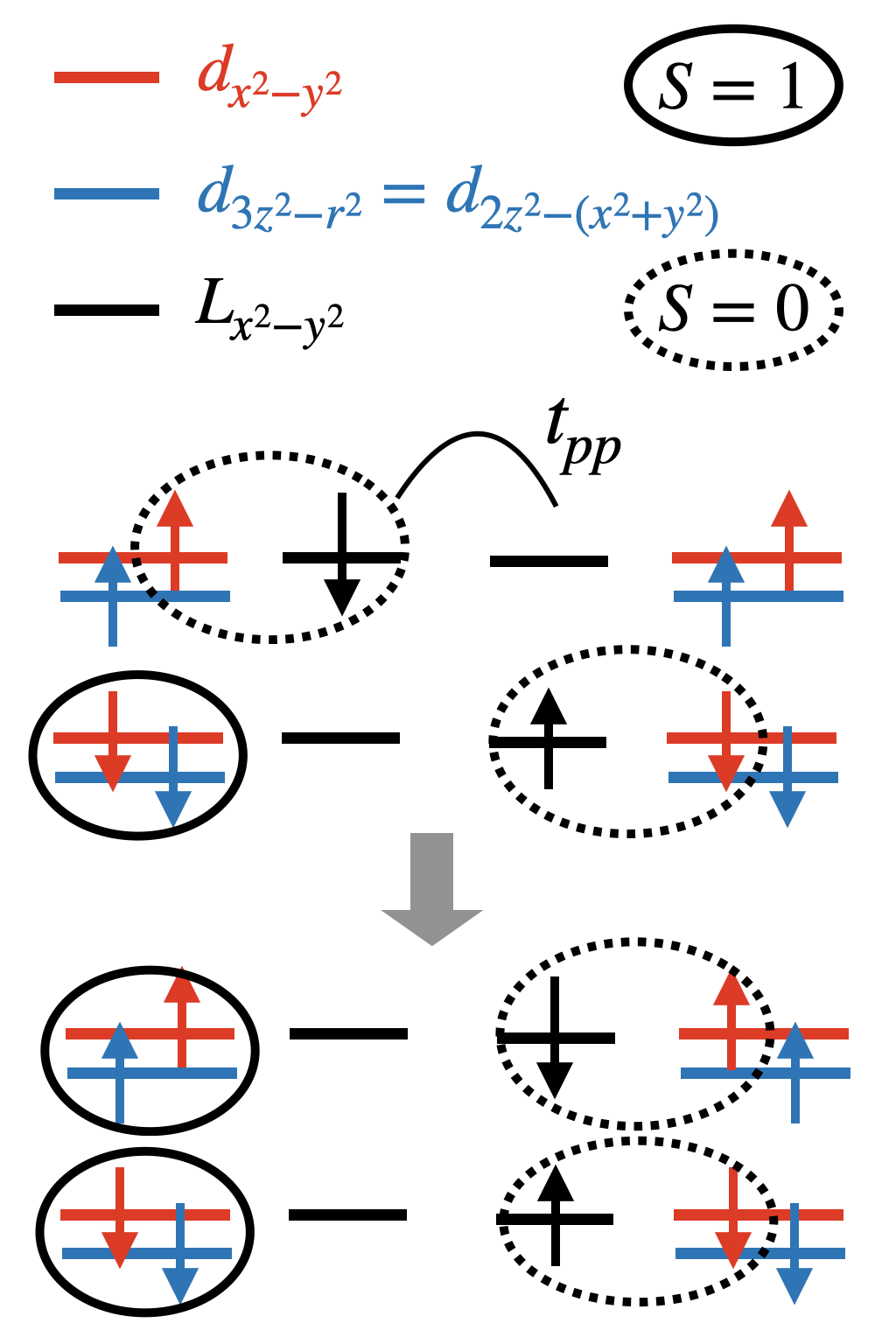,width=.4\textwidth, clip}
\caption{Schematic picture \textcolor{black}{of two possible spin and charge order phases suggested by our results, corresponding to a checkerboard ordering of  5-5 holes clusters (top panel; only two clusters are shown), and of   4-6 holes clusters (bottom panel). The legend at the top shows that  different orbitals are labeled with different colors, and solid/dashed eclipses denote  triplet/singlet ordering of the two orbitals enclosed. Inter-cluster hybridizations and correlations,  suggested by the $t_{pp}$ process, are ignored in our calculations.}}
\label{cdwjpg}
\end{figure}

\textcolor{black}{We now speculate on  possible ordered phases suggested by these cluster results}.
We start with the 5-hole state in which the 5th hole is in $L_{x^2-y^2}$ either in the upper or lower plane. Switching off the inter-planar superexchange between the $d_{z^2}$ orbitals leads to the potential ordering of the $L_{x^2-y^2}$ in a staggered fashion between the upper and lower NiO$_2$ planes.   \textcolor{black}{This suggests a possible ordered phase where each plane would have a checkerboard like ordering of $d^8L$ (spin-$\frac{1}{2}$ from the $d_{3z^2-r^2}$ hole) and $d^8$ (spin-1) respectively, with an out-of-phase ordering between the planes. A picture of this kind of potential ordering is presented in Fig.~\ref{cdwjpg}, whose top panel depicts this ordering represented by an alternation of 3-2 holes for the upper and  2-3 holes in the lower NiO$_2$ planes, which is then repeated in a checkerboard fashion.}
This is reminiscent of the ordering that occurs in the cubic rare-earth nickelates ReNiO$_3$ if we look at the NiO$_2$ plane, {\em i.e.}  the so-called bond or charge disproportionated phase. This out-of-phase charge ordering is intimately intertwined with the spin ordering of 1/2-1 and 1-1/2 in two planes, as shown in the top panel of Fig.~\ref{cdwjpg}. 


\textcolor{black}{An alternative is the in-phase ordered state shown in the bottom panel of Fig.~\ref{cdwjpg}, corresponding to a checkerboard arrangement of  4+6 holes clusters. This also results in intertwined charge and spin ordering, both very different from the ones in the 5+5 ordered phase discussed above. The two layers are still out-of-phase for spin ordering of  1;1 and 1/2;1/2 respectively, while the hole occupation is now the same in both layers, alternating between +2;+2 and +3;+3 in a checkerboard fashion. }

These two ordered states are connected by hopping of the 5th hole between Ni$_2$O$_9$ clusters, as suggested in the top panel. 
We note that this inter-cluster  $t_{pp}$ hopping process links the lowest energy cluster configurations only if the in-plane order of the 1/2-1 (1-1/2) spins in the top (bottom) layer is ferromagnetic, as depicted in the top panel. This is in fact one of the possible magnetic orderings found in our hybrid functional calculations~\cite{Kateryna}.

These considerations suggest that both the 6+4 and 5+5 hole clusters' checkerboard-like configurations would result in intertwined SDW and CDW orders, but of very different character. The 5+5 configuration has a different charge and spin alternation within the Ni$_2$O$_9$ clusters, while the 6+4 one has the charge and spin ordering between the Ni$_2$O$_9$ clusters.

%

\textcolor{black}{We can try to get a very rough estimate for which of these potential CDW-SDW orders are more stable from the energy difference  $\Delta E = E_{6+4}-E_{5+5}$ between the GS energy of a 6- and a 4-hole clusters $E_{6+4} = E_{GS,6}+E_{GS,4}$ and that of two 5-hole clusters, $E_{5+5} = 2E_{GS,5}$. We find this  $\Delta E $ to be negative but with a very small magnitude (10 meV or less at all considered pressure), favouring the 4+6 hole clusters at ambient pressure versus the 5+5 hole clusters at the higher pressures considered. This is an interesting result, however because this energy scale is so small, and because this estimate does not take into account the contributions to $\Delta E$ due to inter-cluster hopping and correlations, we believe that any conclusions based on this value would be too much of a speculation. Bigger clusters or other methods are needed to get more trustworthy estimates.}

\textcolor{black}{Indeed, it is important to note that inter cluster hopping will likely result in strong fluctuations in the local charge and spin between the 5-5 and 6-4 clusters, leading to a smearing out of these local spins and charge densities. This could have a pronounced temperature and pressure dependence, which is a subject of present investigations. }

We anticipate obtaining information regarding the preferred ground state ordering by studying larger clusters with reduced Hilbert spaces. Further investigation of this problem is underway by including the inter-cluster interactions both regarding one particle hopping as well as the exchange interactions via virtual hopping processes.

\section{Summary and outlook}
Our first conclusion is that for parameters close to the DFT based values, the dominant configurations contributing to the ground state do not involve a hole on the apical O between the two NiO$_2$ layers. This is the case both for the average occupation of 5-holes per  Ni$_2$O$_9$ cluster, but also for the charge fluctuations involving 6 and 4 holes states per Ni pair.
Specifically, it is found that the ground state of 5-holes average occupation cluster involves the states $d^8L$-$d^8$ and $d^8$-$d^8L$ with the ligand hole with $x^2-y^2$ symmetry located in either the upper or the lower plane, and forming a Zhang-Rice like singlet state with the $d_{x^2-y^2}$ orbital. This  leaves a spin-1/2 from the hole occupying the $d_{3z^2-r^2}$ orbital of that Ni,  and a spin-1 $d^8$ state for the Ni in the other plane. These two spins couple via a superexchange into a total spin of 1/2 or 3/2, separated by a splitting of about 200 meV. There are also very high energy $S=5/2$ states that are about 4\,eV higher in energy.

\textcolor{black}{While the appearance of Zhang-Rice like singlets between the Cu $3d_{x^2-y^2}$ and the in-plane $L_{x^2-y^2}$ is very reminiscent of cuprate physics, there are also very stark differences. First, taken at face values, our results predict that in the undoped bilayer La$_3$Ni$_2$O$_7$,  50$\%$ of all Ni host such a Zhang-Rice singlet like state. This is a density far in excess of that in even extremely overdoped cuprates. Second, this system also has the additional spins, either 1/2 from the  $3d_{3z^2-r^2}$ orbital of a Ni that hosts a Zhang-Rice singlet like state, or $s=1$ for a Ni that does not. This is to be contrasted with the cuprate situation, where Cu that host a  Zhang-Rice singlet have $s=0$, while Cu that do not have $s=1/2$. It is also important to mention that the $C_4$ symmetry of the cluster favours a Zhang-Rice singlet like state. Other symmetries might favor 3-spin polarons, as is also known to be the case in cuprates~\cite{Bayo2011,EmeryReiter}.}

\textcolor{black}{Our results allow us to also speculate on possible intertwinned CDW and SDW orders in a bilayer made of such 5-holes clusters arranged in a checkerboard pattern, if we assume that spin and charge fluctuations due to inter-cluster hybridization and correlations do not completely 'melt' it. We also highlight the very different intertwinned CDW and SDW orders in a bilayer of 4-holes and 6-holes clusters  arranged in a checkerboard pattern. Interestingly, similar charge and spin distributions  have been recently measured with neutrons and $\mu$SR~\cite{plokhikh_unraveling_2025,yashima_microscopic_2025} although arranged in a larger unit cell comprising 4 clusters. This is why  investigations of various multi-cluster configurations relevant to a more realistic modeling of La$_3$Ni$_2$O$_7$ are necessary before reliable conclusions can be drawn. We are actively engaged in further investigations of this kind, however we believe that this work already provides valuable information, based on reliable values for the various exchange interactions, regarding the nature of the low-energy  charge and spin densities together with the most important configurations that should be included in any cluster or DMFT type of study of this material.}

\section{ACKNOWLEDGMENTS}
\textcolor{black}{Guiwen Jiang and Chenye Qin contributed equally.} We would like to thank Karsten Held for illuminating discussions.
G.~J.~, C.~Q.~, and M.~J.~acknowledge the support by National Natural Science Foundation of China (NSFC) Grant No.~12174278, startup fund from Soochow University, and Priority Academic Program Development (PAPD) of Jiangsu Higher Education Institutions. 
L.~S.~acknowledges support from the National Natural Science Foundation of China (Grant No.~12422407).
K.~F., M.~B. and G.~A.~S. are funded by the Quantum Matter Institute (QMI) at University of British Columbia and by the Natural Sciences and Engineering Research Council of Canada (NSERC).
Calculations have been done mainly on the Soochow University and the national supercomputing center Xi'an at Northwest University.

\bibliographystyle{apsrev4-1}
\bibliographystyle{unsrt}
\bibliography{main}

\begin{thebibliography}{10}

\bibitem{li2019superconductivity}
Danfeng Li, Kyuho Lee, Bai~Yang Wang, Motoki Osada, Samuel Crossley, Hye~Ryoung
  Lee, Yi~Cui, Yasuyuki Hikita, and Harold~Y Hwang.
\newblock Superconductivity in an infinite-layer nickelate.
\newblock {\em Nature}, 572(7771):624--627, 2019.

\bibitem{Li2020}
Danfeng Li, Bai~Yang Wang, Kyuho Lee, Shannon~P. Harvey, Motoki Osada, Berit~H.
  Goodge, Lena~F. Kourkoutis, and Harold~Y. Hwang.
\newblock Superconducting dome in
  {${\mathrm{Nd}}_{1\ensuremath{-}x}{\mathrm{Sr}}_{x}{\mathrm{NiO}}_{2}$}
  infinite layer films, arxiv:2003.08506.
\newblock {\em Phys. Rev. Lett.}, 125:027001, Jul 2020.

\bibitem{PhysRevLett.125.147003}
Shengwei Zeng, Chi~Sin Tang, Xinmao Yin, Changjian Li, Mengsha Li, Zhen Huang,
  Junxiong Hu, Wei Liu, Ganesh~Ji Omar, Hariom Jani, Zhi~Shiuh Lim, Kun Han,
  Dongyang Wan, Ping Yang, Stephen~John Pennycook, Andrew T.~S. Wee, and
  Ariando Ariando.
\newblock Phase diagram and superconducting dome of infinite-layer
  ${\mathrm{nd}}_{1\ensuremath{-}x}{\mathrm{sr}}_{x}{\mathrm{nio}}_{2}$ thin
  films.
\newblock {\em Phys. Rev. Lett.}, 125:147003, Oct 2020.

\bibitem{gu2020single}
Qiangqiang Gu, Yueying Li, Siyuan Wan, Huazhou Li, Wei Guo, Huan Yang, Qing Li,
  Xiyu Zhu, Xiaoqing Pan, Yuefeng Nie, et~al.
\newblock Single particle tunneling spectrum of superconducting nd1-xsrxnio2
  thin films.
\newblock {\em Nature communications}, 11(1):6027, 2020.

\bibitem{Bednorz1986}
J.~G. Bednorz and K.~A. M\"uller.
\newblock Possible high t$_{\rm c}$ superconductivity in the ba--la--cu--o
  system.
\newblock {\em Zeitschrift f\"ur Physik B Condensed Matter}, 64:189--193, June
  1986.

\bibitem{Kitatani2020}
Motoharu {Kitatani}, Liang {Si}, Oleg {Janson}, Ryotaro {Arita}, Zhicheng
  {Zhong}, and Karsten {Held}.
\newblock {Nickelate superconductors -- a renaissance of the one-band Hubbard
  model, arXiv:2002.12230}.
\newblock {\em npj Quantum Materials}, 5:59, 2020.

\bibitem{PhysRevB.109.235126}
Paul Worm, Qisi Wang, Motoharu Kitatani, Izabela Bia\l{}o, Qiang Gao, Xiaolin
  Ren, Jaewon Choi, Diana Csontosov\'a, Ke-Jin Zhou, Xingjiang Zhou, Zhihai
  Zhu, Liang Si, Johan Chang, Jan~M. Tomczak, and Karsten Held.
\newblock Spin fluctuations sufficient to mediate superconductivity in
  nickelates.
\newblock {\em Phys. Rev. B}, 109:235126, Jun 2024.

\bibitem{Nomura2019}
Yusuke Nomura, Motoaki Hirayama, Terumasa Tadano, Yoshihide Yoshimoto, Kazuma
  Nakamura, and Ryotaro Arita.
\newblock Formation of a two-dimensional single-component correlated electron
  system and band engineering in the nickelate superconductor
  {${\mathrm{NdNiO}}_{2}$}.
\newblock {\em Phys. Rev. B}, 100:205138, Nov 2019.

\bibitem{Karp2020}
Jonathan Karp, Antia~S. Botana, Michael~R. Norman, Hyowon Park, Manuel Zingl,
  and Andrew Millis.
\newblock Many-body electronic structure of {${\mathrm{NdNiO}}_{2}$} and
  {${\mathrm{CaCuO}}_{2}$}.
\newblock {\em Phys. Rev. X}, 10:021061, Jun 2020.

\bibitem{Zhang2019}
Guang-Ming Zhang, Yi-feng Yang, and Fu-Chun Zhang.
\newblock Self-doped mott insulator for parent compounds of nickelate
  superconductors.
\newblock {\em Phys. Rev. B}, 101:020501, Jan 2020.

\bibitem{chen2023electronic}
Hanghui Chen, Yi-feng Yang, Guang-Ming Zhang, and Hongquan Liu.
\newblock An electronic origin of charge order in infinite-layer nickelates.
\newblock {\em Nature Communications}, 14(1):5477, 2023.

\bibitem{Lechermann2020}
Frank Lechermann.
\newblock Multiorbital processes rule the
  {${\mathrm{Nd}}_{1\ensuremath{-}x}{\mathrm{Sr}}_{x}{\mathrm{NiO}}_{2}$}
  normal state.
\newblock {\em Phys. Rev. X}, 10:041002, Oct 2020.

\bibitem{PhysRevLett.129.077002}
Andreas Kreisel, Brian~M. Andersen, Astrid~T. R\o{}mer, Ilya~M. Eremin, and
  Frank Lechermann.
\newblock Superconducting instabilities in strongly correlated infinite-layer
  nickelates.
\newblock {\em Phys. Rev. Lett.}, 129:077002, Aug 2022.

\bibitem{Mi2020}
Mi~Jiang, Mona Berciu, and George~A. Sawatzky.
\newblock Critical nature of the ni spin state in doped ${\mathrm{ndnio}}_{2}$.
\newblock {\em Phys. Rev. Lett.}, 124:207004, May 2020.

\bibitem{PhysRevB.102.220501}
Zhan Wang, Guang-Ming Zhang, Yi-feng Yang, and Fu-Chun Zhang.
\newblock Distinct pairing symmetries of superconductivity in infinite-layer
  nickelates.
\newblock {\em Phys. Rev. B}, 102:220501, Dec 2020.

\bibitem{cheng2024evidence}
Bing Cheng, Di~Cheng, Kyuho Lee, Liang Luo, Zhuoyu Chen, Yonghun Lee, Bai~Yang
  Wang, Martin Mootz, Ilias~E Perakis, Zhi-Xun Shen, et~al.
\newblock Evidence for d-wave superconductivity of infinite-layer nickelates
  from low-energy electrodynamics.
\newblock {\em Nature Materials}, 23(6):775--781, 2024.

\bibitem{sun2023signatures}
Hualei Sun, Mengwu Huo, Xunwu Hu, Jingyuan Li, Zengjia Liu, Yifeng Han, Lingyun
  Tang, Zhongquan Mao, Pengtao Yang, Bosen Wang, et~al.
\newblock Signatures of superconductivity near 80 k in a nickelate under high
  pressure.
\newblock {\em Nature}, 621(7979):493--498, 2023.

\bibitem{PhysRevX.14.011040}
G.~Wang, N.~N. Wang, X.~L. Shen, J.~Hou, L.~Ma, L.~F. Shi, Z.~A. Ren, Y.~D. Gu,
  H.~M. Ma, P.~T. Yang, Z.~Y. Liu, H.~Z. Guo, J.~P. Sun, G.~M. Zhang,
  S.~Calder, J.-Q. Yan, B.~S. Wang, Y.~Uwatoko, and J.-G. Cheng.
\newblock Pressure-induced superconductivity in polycrystalline
  ${\mathrm{la}}_{3}{\mathrm{ni}}_{2}{\mathrm{o}}_{7\ensuremath{-}\ensuremath{\delta}}$.
\newblock {\em Phys. Rev. X}, 14:011040, Mar 2024.

\bibitem{yang2024orbital}
Jiangang Yang, Hualei Sun, Xunwu Hu, Yuyang Xie, Taimin Miao, Hailan Luo, Hao
  Chen, Bo~Liang, Wenpei Zhu, Gexing Qu, et~al.
\newblock Orbital-dependent electron correlation in double-layer nickelate
  la3ni2o7.
\newblock {\em Nature Communications}, 15(1):4373, 2024.

\bibitem{hou2023emergence}
Jun Hou, Peng-Tao Yang, Zi-Yi Liu, Jing-Yuan Li, Peng-Fei Shan, Liang Ma, Gang
  Wang, Ning-Ning Wang, Hai-Zhong Guo, Jian-Ping Sun, et~al.
\newblock Emergence of high-temperature superconducting phase in pressurized
  la3ni2o7 crystals.
\newblock {\em Chinese Physics Letters}, 40(11):117302, 2023.

\bibitem{zhou2024evidence}
Yazhou Zhou, Jing Guo, Shu Cai, Hualei Sun, Pengyu Wang, Jinyu Zhao, Jinyu Han,
  Xintian Chen, Yongjin Chen, Qi~Wu, Yang Ding, Tao Xiang, Ho~kwang Mao, and
  Liling Sun.
\newblock Evidence of filamentary superconductivity in pressurized la3ni2o7.
\newblock 2024.

\bibitem{PhysRevLett.132.256503}
Kaiwen Chen, Xiangqi Liu, Jiachen Jiao, Muyuan Zou, Chengyu Jiang, Xin Li,
  Yixuan Luo, Qiong Wu, Ningyuan Zhang, Yanfeng Guo, and Lei Shu.
\newblock Evidence of spin density waves in
  ${\mathrm{la}}_{3}{\mathrm{ni}}_{2}{\mathrm{o}}_{7\ensuremath{-}\ensuremath{\delta}}$.
\newblock {\em Phys. Rev. Lett.}, 132:256503, Jun 2024.

\bibitem{jiang2024high}
Kun Jiang, Ziqiang Wang, and Fu-Chun Zhang.
\newblock High-temperature superconductivity in la3ni2o7.
\newblock {\em Chinese Physics Letters}, 41(1):017402, 2024.

\bibitem{wang2024normal}
Meng Wang, Hai-Hu Wen, Tao Wu, Dao-Xin Yao, and Tao Xiang.
\newblock Normal and superconducting properties of la$_3$ni$_2$o$_7$.
\newblock 2024.

\bibitem{dan2024spindensitywave}
Zhao Dan, Yanbing Zhou, Mengwu Huo, Yu~Wang, Linpeng Nie, Meng Wang, Tao Wu,
  and Xianhui Chen.
\newblock Spin-density-wave transition in double-layer nickelate la3ni2o7.
\newblock 2024.

\bibitem{wu2024superexchange}
W{\'e}i W{\'u}, Zhihui Luo, Dao-Xin Yao, and Meng Wang.
\newblock Superexchange and charge transfer in the nickelate superconductor
  la3ni2o7 under pressure.
\newblock {\em Science China Physics, Mechanics \& Astronomy}, 67(11):117402,
  2024.

\bibitem{xie2024neutron}
Tao Xie, Mengwu Huo, Xiaosheng Ni, Feiran Shen, Xing Huang, Hualei Sun,
  Helen~C. Walker, Devashibhai Adroja, Dehong Yu, Bing Shen, Lunhua He, Kun
  Cao, and Meng Wang.
\newblock Neutron scattering studies on the high-$t_c$ superconductor
  la$_3$ni$_2$o$_{7-\delta}$ at ambient pressure.
\newblock 2024.

\bibitem{yi2024antiferromagnetic}
Xin-Wei Yi, Ying Meng, Jia-Wen Li, Zheng-Wei Liao, Jing-Yang You, Bo~Gu, and
  Gang Su.
\newblock Antiferromagnetic ground state, charge density waves and oxygen
  vacancies induced metal-insulator transition in pressurized
  {La}$_{3}${Ni}$_{2}${O}$_{7}$.
\newblock 2024.

\bibitem{PhysRevB.108.L201121}
Frank Lechermann, Jannik Gondolf, Steffen B\"otzel, and Ilya~M. Eremin.
\newblock Electronic correlations and superconducting instability in
  ${\mathrm{la}}_{3}{\mathrm{ni}}_{2}{\mathrm{o}}_{7}$ under high pressure.
\newblock {\em Phys. Rev. B}, 108:L201121, Nov 2023.

\bibitem{wang2024electronic}
Yuxin Wang, Kun Jiang, Ziqiang Wang, Fu-Chun Zhang, and Jiangping Hu.
\newblock Electronic structure and superconductivity in bilayer
  la$_3$ni$_2$o$_7$.
\newblock 2024.

\bibitem{PhysRevB.108.125105}
D.~A. Shilenko and I.~V. Leonov.
\newblock Correlated electronic structure, orbital-selective behavior, and
  magnetic correlations in double-layer
  ${\mathrm{la}}_{3}{\mathrm{ni}}_{2}{\mathrm{o}}_{7}$ under pressure.
\newblock {\em Phys. Rev. B}, 108:125105, Sep 2023.

\bibitem{cui2024strain}
Ting Cui, Songhee Choi, Ting Lin, Chen Liu, Gang Wang, Ningning Wang, Shengru
  Chen, Haitao Hong, Dongke Rong, Qianying Wang, et~al.
\newblock Strain-mediated phase crossover in ruddlesden--popper nickelates.
\newblock {\em Communications Materials}, 5(1):32, 2024.

\bibitem{abadi2024electronic}
Sebastien~N. Abadi, Ke-Jun Xu, Eder~G. Lomeli, Pascal Puphal, Masahiko Isobe,
  Yong Zhong, Alexei~V. Fedorov, Sung-Kwan Mo, Makoto Hashimoto, Dong-Hui Lu,
  Brian Moritz, Bernhard Keimer, Thomas~P. Devereaux, Matthias Hepting, and
  Zhi-Xun Shen.
\newblock Electronic structure of the alternating monolayer-trilayer phase of
  la3ni2o7.
\newblock 2024.

\bibitem{chen2024polymorphism}
Xinglong Chen, Junjie Zhang, Arashdeep~S Thind, Shekhar Sharma, Harrison
  LaBollita, Gordon Peterson, Hong Zheng, Daniel~P Phelan, Antia~S Botana,
  Robert~F Klie, et~al.
\newblock Polymorphism in the ruddlesden--popper nickelate la3ni2o7: Discovery
  of a hidden phase with distinctive layer stacking.
\newblock {\em Journal of the American Chemical Society}, 2024.

\bibitem{wang2024long}
Haozhe Wang, Long Chen, Aya Rutherford, Haidong Zhou, and Weiwei Xie.
\newblock Long-range structural order in a hidden phase of ruddlesden--popper
  bilayer nickelate la3ni2o7.
\newblock {\em Inorganic Chemistry}, 63(11):5020--5026, 2024.

\bibitem{zhang_high-temperature_2024}
Yanan Zhang, Dajun Su, Yanen Huang, Zhaoyang Shan, Hualei Sun, Mengwu Huo,
  Kaixin Ye, Jiawen Zhang, Zihan Yang, Yongkang Xu, Yi~Su, Rui Li, Michael
  Smidman, Meng Wang, Lin Jiao, and Huiqiu Yuan.
\newblock High-temperature superconductivity with zero resistance and
  strange-metal behaviour in {La3Ni2O7}.
\newblock {\em Nature Physics}, pages 1--5, June 2024.

\bibitem{La2PrNi2O7_2024}
Ningning Wang, Gang Wang, Xiaoling Shen, Jun Hou, Jun Luo, Xiaoping Ma, Huaixin
  Yang, Lifen Shi, Jie Dou, Jie Feng, Jie Yang, Yunqing Shi, Zhian Ren, Hanming
  Ma, Pengtao Yang, Ziyi Liu, Yue Liu, Hua Zhang, Xiaoli Dong, Yuxin Wang, Kun
  Jiang, Jiangping Hu, Shoko Nagasaki, Kentaro Kitagawa, Stuart Calder,
  Jiaqiang Yan, Jianping Sun, Bosen Wang, Rui Zhou, Yoshiya Uwatoko, and
  Jinguang Cheng.
\newblock Bulk high-temperature superconductivity in pressurized tetragonal
  la2prni2o7.
\newblock {\em Nature}, 2024.

\bibitem{dong_visualization_2024}
Zehao Dong, Mengwu Huo, Jie Li, Jingyuan Li, Pengcheng Li, Hualei Sun, Lin Gu,
  Yi~Lu, Meng Wang, Yayu Wang, and Zhen Chen.
\newblock Visualization of oxygen vacancies and self-doped ligand holes in
  {La3Ni2O7}.
\newblock {\em Nature}, 630(8018):847--852, June 2024.

\bibitem{Cava24}
Ran Gao, Lun Jin, Shuyuan Huyan, Danrui Ni, Haozhe Wang, Xianghan Xu, Sergey~L.
  Bud'ko, Paul Canfield, Weiwei Xie, and Robert~J. Cava.
\newblock Is la3ni2o6.5 a bulk superconducting nickelate?
\newblock {\em ACS Applied Materials \& Interfaces}, 16(49):66857--66864, 12
  2024.

\bibitem{PhysRevB.108.214522}
Zhiguang Liao, Lei Chen, Guijing Duan, Yiming Wang, Changle Liu, Rong Yu, and
  Qimiao Si.
\newblock Electron correlations and superconductivity in
  ${\mathrm{la}}_{3}{\mathrm{ni}}_{2}{\mathrm{o}}_{7}$ under pressure tuning.
\newblock {\em Phys. Rev. B}, 108:214522, Dec 2023.

\bibitem{PhysRevB.109.L180502}
Steffen B\"otzel, Frank Lechermann, Jannik Gondolf, and Ilya~M. Eremin.
\newblock Theory of magnetic excitations in the multilayer nickelate
  superconductor ${\mathrm{la}}_{3}{\mathrm{ni}}_{2}{\mathrm{o}}_{7}$.
\newblock {\em Phys. Rev. B}, 109:L180502, May 2024.

\bibitem{PhysRevMaterials.2.125001}
Yasuhide Mochizuki, Hirofumi Akamatsu, Yu~Kumagai, and Fumiyasu Oba.
\newblock Strain-engineered peierls instability in layered perovskite
  ${\mathrm{la}}_{3}{\mathrm{ni}}_{2}{\mathrm{o}}_{7}$ from first principles.
\newblock {\em Phys. Rev. Mater.}, 2:125001, Dec 2018.

\bibitem{chen2023critical}
Xuejiao Chen, Peiheng Jiang, Jie Li, Zhicheng Zhong, and Yi~Lu.
\newblock Critical charge and spin instabilities in superconducting
  la$_3$ni$_2$o$_7$.
\newblock 2023.

\bibitem{zhan2024cooperation}
Jun Zhan, Yuhao Gu, Xianxin Wu, and Jiangping Hu.
\newblock Cooperation between electron-phonon coupling and electronic
  interaction in bilayer nickelates la$_3$ni$_2$o$_7$.
\newblock 2024.

\bibitem{PhysRevB.108.L140505}
Qing-Geng Yang, Da~Wang, and Qiang-Hua Wang.
\newblock Possible ${s}_{\ifmmode\pm\else\textpm\fi{}}$-wave superconductivity
  in ${\mathrm{la}}_{3}{\mathrm{ni}}_{2}{\mathrm{o}}_{7}$.
\newblock {\em Phys. Rev. B}, 108:L140505, Oct 2023.

\bibitem{PhysRevB.109.L220506}
Qing-Geng Yang, Kai-Yue Jiang, Da~Wang, Hong-Yan Lu, and Qiang-Hua Wang.
\newblock Effective model and ${s}_{\ifmmode\pm\else\textpm\fi{}}$-wave
  superconductivity in trilayer nickelate
  ${\mathrm{la}}_{4}{\mathrm{ni}}_{3}{\mathrm{o}}_{10}$.
\newblock {\em Phys. Rev. B}, 109:L220506, Jun 2024.

\bibitem{luo2024high}
Zhihui Luo, Biao Lv, Meng Wang, W{\'e}i W{\'u}, and Dao-Xin Yao.
\newblock High-t c superconductivity in {L}a3{N}i2{O}7 based on the bilayer
  two-orbital tj model.
\newblock {\em NPJ Quantum Materials}, 9(1):61, 2024.

\bibitem{ouyang2024absence}
Zhenfeng Ouyang, Miao Gao, and Zhong-Yi Lu.
\newblock Absence of electron-phonon coupling superconductivity in the bilayer
  phase of la3ni2o7 under pressure.
\newblock {\em npj Quantum Materials}, 9(1):80, 2024.

\bibitem{PhysRevB.109.104508}
Griffin Heier, Kyungwha Park, and Sergey~Y. Savrasov.
\newblock Competing ${d}_{xy}$ and ${s}_{\ifmmode\pm\else\textpm\fi{}}$ pairing
  symmetries in superconducting
  ${\mathrm{la}}_{3}{\mathrm{ni}}_{2}{\mathrm{o}}_{7}$:
  $\mathrm{LDA}+\mathrm{FLEX}$ calculations.
\newblock {\em Phys. Rev. B}, 109:104508, Mar 2024.

\bibitem{PhysRevLett.132.146002}
Chen Lu, Zhiming Pan, Fan Yang, and Congjun Wu.
\newblock Interlayer-coupling-driven high-temperature superconductivity in
  ${\mathrm{la}}_{3}{\mathrm{ni}}_{2}{\mathrm{o}}_{7}$ under pressure.
\newblock {\em Phys. Rev. Lett.}, 132:146002, Apr 2024.

\bibitem{PhysRevB.108.165141}
Yang Zhang, Ling-Fang Lin, Adriana Moreo, Thomas~A. Maier, and Elbio Dagotto.
\newblock Trends in electronic structures and
  ${s}_{\ifmmode\pm\else\textpm\fi{}}$-wave pairing for the rare-earth series
  in bilayer nickelate superconductor
  ${R}_{3}{\mathrm{ni}}_{2}{\mathrm{o}}_{7}$.
\newblock {\em Phys. Rev. B}, 108:165141, Oct 2023.

\bibitem{PhysRevB.109.165154}
Yi-Heng Tian, Yin Chen, Jia-Ming Wang, Rong-Qiang He, and Zhong-Yi Lu.
\newblock Correlation effects and concomitant two-orbital
  ${s}_{\ifmmode\pm\else\textpm\fi{}}$-wave superconductivity in
  ${\mathrm{la}}_{3}{\mathrm{ni}}_{2}{\mathrm{o}}_{7}$ under high pressure.
\newblock {\em Phys. Rev. B}, 109:165154, Apr 2024.

\bibitem{PhysRevLett.133.096002}
Siheon Ryee, Niklas Witt, and Tim~O. Wehling.
\newblock Quenched pair breaking by interlayer correlations as a key to
  superconductivity in ${\mathrm{la}}_{3}{\mathrm{ni}}_{2}{\mathrm{o}}_{7}$.
\newblock {\em Phys. Rev. Lett.}, 133:096002, Aug 2024.

\bibitem{PhysRevLett.132.036502}
Xing-Zhou Qu, Dai-Wei Qu, Jialin Chen, Congjun Wu, Fan Yang, Wei Li, and Gang
  Su.
\newblock Bilayer
  ${t\text{\ensuremath{-}}J\text{\ensuremath{-}}J}_{\ensuremath{\perp}}$ model
  and magnetically mediated pairing in the pressurized nickelate
  ${\mathrm{la}}_{3}{\mathrm{ni}}_{2}{\mathrm{o}}_{7}$.
\newblock {\em Phys. Rev. Lett.}, 132:036502, Jan 2024.

\bibitem{PhysRevB.108.174501}
Junkang Huang, Z.~D. Wang, and Tao Zhou.
\newblock Impurity and vortex states in the bilayer high-temperature
  superconductor ${\mathrm{la}}_{3}{\mathrm{ni}}_{2}{\mathrm{o}}_{7}$.
\newblock {\em Phys. Rev. B}, 108:174501, Nov 2023.

\bibitem{PhysRevLett.133.136001}
Yang Zhang, Ling-Fang Lin, Adriana Moreo, Thomas~A. Maier, and Elbio Dagotto.
\newblock Prediction of ${s}^{\ifmmode\pm\else\textpm\fi{}}$-wave
  superconductivity enhanced by electronic doping in trilayer nickelates
  ${\mathrm{la}}_{4}{\mathrm{ni}}_{3}{\mathrm{o}}_{10}$ under pressure.
\newblock {\em Phys. Rev. Lett.}, 133:136001, Sep 2024.

\bibitem{PhysRevB.108.L140504}
Qiong Qin and Yi-feng Yang.
\newblock High-${T}_{c}$ superconductivity by mobilizing local spin singlets
  and possible route to higher ${T}_{c}$ in pressurized
  ${\mathrm{la}}_{3}{\mathrm{ni}}_{2}{\mathrm{o}}_{7}$.
\newblock {\em Phys. Rev. B}, 108:L140504, Oct 2023.

\bibitem{PhysRevB.110.L060510}
Yang Zhang, Ling-Fang Lin, Adriana Moreo, Thomas~A. Maier, and Elbio Dagotto.
\newblock Electronic structure, self-doping, and superconducting instability in
  the alternating single-layer trilayer stacking nickelates
  ${\mathrm{la}}_{3}{\mathrm{ni}}_{2}{\mathrm{o}}_{7}$.
\newblock {\em Phys. Rev. B}, 110:L060510, Aug 2024.

\bibitem{geisler2024optical}
Benjamin Geisler, Laura Fanfarillo, James~J Hamlin, Gregory~R Stewart,
  Richard~G Hennig, and PJ~Hirschfeld.
\newblock Optical properties and electronic correlations in la3ni2o7 bilayer
  nickelates under high pressure.
\newblock {\em npj Quantum Materials}, 9(1):89, 2024.

\bibitem{zhang2024doping}
Haiyang Zhang, yujie Bai, Fan-Jie Kong, Xiuqiang Wu, Yuheng Xing, and Ning Xu.
\newblock Doping evolution of the normal state magneticexcitations in
  pressurized la3ni2o7.
\newblock {\em New Journal of Physics}, 2024.

\bibitem{chen2024electronic}
Xiaoyang Chen, Jaewon Choi, Zhicheng Jiang, Jiong Mei, Kun Jiang, Jie Li,
  Stefano Agrestini, Mirian Garcia-Fernandez, Hualei Sun, Xing Huang, et~al.
\newblock Electronic and magnetic excitations in la3ni2o7.
\newblock {\em Nature communications}, 15(1):9597, 2024.

\bibitem{PhysRevMaterials.8.L111801}
Harrison LaBollita, Victor Pardo, Michael~R. Norman, and Antia~S. Botana.
\newblock Assessing spin-density wave formation in
  ${\mathrm{la}}_{3}{\mathrm{ni}}_{2}{\mathrm{o}}_{7}$ from electronic
  structure calculations.
\newblock {\em Phys. Rev. Mater.}, 8:L111801, Nov 2024.

\bibitem{kakoi2024multiband}
Masataka Kakoi, Takashi Oi, Yujiro Ohshita, Mitsuharu Yashima, Kazuhiko Kuroki,
  Takeru Kato, Hidefumi Takahashi, Shintaro Ishiwata, Yoshinobu Adachi, Naoyuki
  Hatada, et~al.
\newblock Multiband metallic ground state in multilayered nickelates la3ni2o7
  and la4ni3o10 probed by 139la-nmr at ambient pressure.
\newblock {\em Journal of the Physical Society of Japan}, 93(5):053702, 2024.

\bibitem{PhysRevLett.131.206501}
Viktor Christiansson, Francesco Petocchi, and Philipp Werner.
\newblock Correlated electronic structure of
  ${\mathrm{la}}_{3}{\text{ni}}_{2}{\mathrm{o}}_{7}$ under pressure.
\newblock {\em Phys. Rev. Lett.}, 131:206501, Nov 2023.

\bibitem{PhysRevB.110.195135}
Ling-Fang Lin, Yang Zhang, Nitin Kaushal, Gonzalo Alvarez, Thomas~A. Maier,
  Adriana Moreo, and Elbio Dagotto.
\newblock Magnetic phase diagram of a two-orbital model for bilayer nickelates
  with varying doping.
\newblock {\em Phys. Rev. B}, 110:195135, Nov 2024.

\bibitem{PhysRevB.110.L140508}
Xin-Wei Yi, Ying Meng, Jia-Wen Li, Zheng-Wei Liao, Wei Li, Jing-Yang You,
  Bo~Gu, and Gang Su.
\newblock Nature of charge density waves and metal-insulator transition in
  pressurized ${\mathrm{la}}_{3}{\mathrm{ni}}_{2}{\mathrm{o}}_{7}$.
\newblock {\em Phys. Rev. B}, 110:L140508, Oct 2024.

\bibitem{shen2023effective}
Yang Shen, Mingpu Qin, and Guang-Ming Zhang.
\newblock Effective bi-layer model hamiltonian and density-matrix
  renormalization group study for the high-t c superconductivity in la3ni2o7
  under high pressure.
\newblock {\em Chinese Physics Letters}, 40(12):127401, 2023.

\bibitem{ouyang2024absencephononmediatedsuperconductivityla3ni2o7}
Zhenfeng Ouyang, Miao Gao, and Zhong-Yi Lu.
\newblock Absence of phonon-mediated superconductivity in la$_3$ni$_2$o$_7$
  under pressure.
\newblock 2024.

\bibitem{PhysRev.136.B864}
P.~Hohenberg and W.~Kohn.
\newblock Inhomogeneous electron gas.
\newblock {\em Phys. Rev.}, 136:B864--B871, Nov 1964.

\bibitem{PhysRev.140.A1133}
W.~Kohn and L.~J. Sham.
\newblock Self-consistent equations including exchange and correlation effects.
\newblock {\em Phys. Rev.}, 140:A1133--A1138, Nov 1965.

\bibitem{Kateryna}
Ilya~Elfimov Kateryna~Foyevtsova and George~A. Sawatzky.
\newblock {\em to be published}.

\bibitem{kresse1996efficiency}
Georg Kresse and J{\"u}rgen Furthm{\"u}ller.
\newblock Efficiency of ab-initio total energy calculations for metals and
  semiconductors using a plane-wave basis set.
\newblock {\em Computational materials science}, 6(1):15--50, 1996.

\bibitem{PhysRevB.54.11169}
G.~Kresse and J.~Furthm\"uller.
\newblock Efficient iterative schemes for ab initio total-energy calculations
  using a plane-wave basis set.
\newblock {\em Phys. Rev. B}, 54:11169--11186, Oct 1996.

\bibitem{blaha2001wien2k}
Peter Blaha, Karlheinz Schwarz, GKH Madsen, Dieter Kvasnicka, and Joachim
  Luitz.
\newblock wien2k,an augmented plane wave+ local orbitals program for
  calculating crystal properties.
\newblock {\em Wien2k,an augmented plane wave+ local orbitals program for
  calculating crystal properties}, 2001.

\bibitem{Schwarz2002}
K.~Schwarz, P.~Blaha, and G.~K.~H. Madsen.
\newblock Electronic structure calculations of solids using the wien2k package
  for material sciences.
\newblock {\em Comp. Phys. Comm.}, 147:71 -- 76, 2002.

\bibitem{PhysRevLett.77.3865}
John~P. Perdew, Kieron Burke, and Matthias Ernzerhof.
\newblock Generalized gradient approximation made simple.
\newblock {\em Phys. Rev. Lett.}, 77:3865--3868, Oct 1996.

\bibitem{PhysRev.52.191}
Gregory~H. Wannier.
\newblock The structure of electronic excitation levels in insulating crystals.
\newblock {\em Phys. Rev.}, 52:191--197, Aug 1937.

\bibitem{mostofi2008wannier90}
Arash~A. Mostofi, Jonathan~R. Yates, Young-Su Lee, Ivo Souza, David Vanderbilt,
  and Nicola Marzari.
\newblock wannier90: A tool for obtaining maximally-localised wannier
  functions.
\newblock {\em Computer Physics Communications}, 178(9):685--699, 2008.

\bibitem{RevModPhys.84.1419}
Nicola Marzari, Arash~A. Mostofi, Jonathan~R. Yates, Ivo Souza, and David
  Vanderbilt.
\newblock Maximally localized wannier functions: Theory and applications.
\newblock {\em Rev. Mod. Phys.}, 84:1419--1475, Oct 2012.

\bibitem{kunevs2010wien2wannier}
Jan Kuneš, Ryotaro Arita, Philipp Wissgott, Alessandro Toschi, Hiroaki Ikeda,
  and Karsten Held.
\newblock Wien2wannier: From linearized augmented plane waves to maximally
  localized wannier functions.
\newblock {\em Computer Physics Communications}, 181(11):1888--1895, 2010.

\bibitem{Mi2020a}
Mi~Jiang, Mirko Moeller, Mona Berciu, and George~A. Sawatzky.
\newblock Relevance of $\mathrm{Cu}--3d$ multiplet structure in models of
  high-${T}_{c}$ cuprates.
\newblock {\em Phys. Rev. B}, 101:035151, Jan 2020.

\bibitem{Mi2022}
Mi~Jiang, Mona Berciu, and George~A. Sawatzky.
\newblock Stabilization of singlet hole-doped state in infinite-layer nickelate
  superconductors.
\newblock {\em Phys. Rev. B}, 106:115150, Sep 2022.

\bibitem{Mi2023}
Chenye Qin and Mi~Jiang.
\newblock Inversion symmetry breaking in bilayer multi-orbital hubbard model
  with impurity approximation.
\newblock 2023.

\bibitem{PhysRevB.108.155147}
Chenye Qin, Mi~Jiang, and Liang Si.
\newblock Effects of different concentrations of topotactic hydrogen impurities
  on the electronic structure of nickelate superconductors.
\newblock {\em Phys. Rev. B}, 108:155147, Oct 2023.

\bibitem{bisogni_ground-state_2016}
Valentina Bisogni, Sara Catalano, Robert~J. Green, Marta Gibert, Raoul
  Scherwitzl, Yaobo Huang, Vladimir~N. Strocov, Pavlo Zubko, Shadi Balandeh,
  Jean-Marc Triscone, George Sawatzky, and Thorsten Schmitt.
\newblock Ground-state oxygen holes and the metal--insulator transition in the
  negative charge-transfer rare-earth nickelates.
\newblock {\em Nature Communications}, 7(1):13017, October 2016.

\bibitem{PhysRevLett.112.106404}
Steve Johnston, Anamitra Mukherjee, Ilya Elfimov, Mona Berciu, and George~A.
  Sawatzky.
\newblock Charge disproportionation without charge transfer in the
  rare-earth-element nickelates as a possible mechanism for the metal-insulator
  transition.
\newblock {\em Phys. Rev. Lett.}, 112:106404, Mar 2014.

\bibitem{PhysRevB.94.195127}
R.~J. Green, M.~W. Haverkort, and G.~A. Sawatzky.
\newblock Bond disproportionation and dynamical charge fluctuations in the
  perovskite rare-earth nickelates.
\newblock {\em Phys. Rev. B}, 94:195127, Nov 2016.

\bibitem{PhysRevLett.100.136406}
John~P. Perdew, Adrienn Ruzsinszky, G\'abor~I. Csonka, Oleg~A. Vydrov,
  Gustavo~E. Scuseria, Lucian~A. Constantin, Xiaolan Zhou, and Kieron Burke.
\newblock Restoring the density-gradient expansion for exchange in solids and
  surfaces.
\newblock {\em Phys. Rev. Lett.}, 100:136406, Apr 2008.

\bibitem{PhysRevB.38.11322}
J.~Ghijsen, L.~H. Tjeng, J.~van Elp, H.~Eskes, J.~Westerink, G.~A. Sawatzky,
  and M.~T. Czyzyk.
\newblock Electronic structure of ${\mathrm{cu}}_{2}$o and cuo.
\newblock {\em Phys. Rev. B}, 38:11322--11330, Dec 1988.

\bibitem{PhysRevLett.126.127401}
Chang-Jong Kang and Gabriel Kotliar.
\newblock Optical properties of the infinite-layer
  ${\mathrm{la}}_{1\ensuremath{-}x}{\mathrm{sr}}_{x}{\mathrm{nio}}_{2}$ and
  hidden hund's physics.
\newblock {\em Phys. Rev. Lett.}, 126:127401, Mar 2021.

\bibitem{PhysRevB.41.288}
H.~Eskes, L.~H. Tjeng, and G.~A. Sawatzky.
\newblock Cluster-model calculation of the electronic structure of cuo: A model
  material for the high-${T}_{c}$ superconductors.
\newblock {\em Phys. Rev. B}, 41:288--299, Jan 1990.

\bibitem{Bayo2011}
Bayo Lau, Mona Berciu, and George~A. Sawatzky.
\newblock High-spin polaron in lightly doped ${\mathrm{cuo}}_{2}$ planes.
\newblock {\em Phys. Rev. Lett.}, 106:036401, Jan 2011.

\bibitem{EmeryReiter}
V.~J. Emery and G.~Reiter.
\newblock Mechanism for high-temperature superconductivity.
\newblock {\em Phys. Rev. B}, 38:4547--4556, Sep 1988.

\bibitem{plokhikh_unraveling_2025}
Igor Plokhikh, Thomas~J. Hicken, Lukas Keller, Vladimir Pomjakushin, Samuel~H.
  Moody, Pascale Foury-Leylekian, Jonas~J. Krieger, Hubertus Luetkens, Zurab
  Guguchia, Rustem Khasanov, and Dariusz~Jakub Gawryluk.
\newblock Unraveling {Spin} {Density} {Wave} {Order} in {Layered} {Nickelates}
  {L}a$_3${N}i$_2${O}$_7$ and {L}a$_2${P}r$_2${N}i$_2${O}$_7$ via {Neutron}
  {Diffraction}.
\newblock March 2025.
\newblock arXiv:2503.05287.

\bibitem{yashima_microscopic_2025}
Mitsuharu Yashima, Nina Seto, Yujiro Oshita, Masataka Kakoi, Hiroya Sakurai,
  Yoshihiko Takano, and Hidekazu Mukuda.
\newblock Microscopic evidence for spin-spinless stripe order with reduced {Ni}
  moments within $ab$ plane for bilayer nickelate {L}a$_3${N}i$_2${O}$_7$
  probed by$^{139}$ {L}a-{NQR}.
\newblock March 2025.
\newblock arXiv:2503.09288.

\end{thebibliography}

\end{document}